\documentclass[british,nofootinbib, 11pt,a4paper]{revtex4-1}
\usepackage[T1]{fontenc}
\usepackage[latin9]{inputenc}
\usepackage{mathrsfs}
\usepackage{mathtools}
\usepackage{amsmath}
\usepackage{amsthm}
\usepackage{amssymb}
\usepackage{esint}

\makeatletter
\theoremstyle{remark}
\newtheorem*{rem*}{\protect\remarkname}

\makeatother

\usepackage{babel}
\providecommand{\remarkname}{Remark}

\begin{document}

\title{An $AdS_{3}$ Dual for Supersymmetric MHV Celestial Amplitudes}

\author{Igor Mol}

\affiliation{State University of Campinas (Unicamp)}
\email{igormol@ime.unicamp.br}

\selectlanguage{british}%
\begin{abstract}
We propose a generalisation of the Wess-Zumino-Novikov-Witten (WZNW)
model, formulated on a holomorphic extension of supersymmetric three-dimensional
Anti-de Sitter $\left(AdS_{3}\right)$ space, which holographically
reproduces the tree-level maximally-helicity-violating (MHV) celestial
amplitudes for gravitons in $\mathcal{N}=8$ supergravity and gluons
in four-dimensional $\mathcal{N}=4$ supersymmetric Yang-Mills (SYM)
theory.
\end{abstract}
\maketitle
\tableofcontents{}

\newpage{}

\section{Introduction}

This study represents the confluence of two distinct lines of inquiry.
The first extends the investigations initiated by \citet{ogawa2024celestial}
and subsequently refined by \citet{mol2024comments} into the correspondence
between celestial conformal field theory (CFT) and (Euclidean) $AdS_{3}$
string theory. The holographic dictionary proposed in \citet{mol2024comments}
exhibited a structural asymmetry in the formulation of celestial vertex
operators for gluons and gravitons within the celestial CFT. Specifically,
while celestial gluon vertex operators were derived exclusively from
solutions to the $AdS_{3}$ string theory equations of motion (whether
from worldsheet primary fields or boundary current algebras on $AdS_{3}$)
the graviton vertex operators relied on an auxiliary set of operators
defined on the celestial sphere. These additional operators effectively
``dressed'' the graviton vertex operators to ensure that their correlators
reproduced the Berends-Giele-Kuijf (BGK) formula describing the tree-level
maximally-helicity-violating (MHV) scattering amplitudes for gravitons
in Einstein's gravity.

In the present work, we derive an alternative representation for celestial
graviton leaf amplitudes, employing an operator factorisation approach
that will be detailed below. It will be demonstrated that, following
the terminology introduced by \citet{cachazo2013gravity}, \citet{cachazo2014gravity}
and \citet{adamo2013twistor}, the graviton leaf amplitudes can be
expressed in terms of multi-graviton wavefunctions defined on a supersymmetric
extension of minitwistor space, $\mathbf{MT}$, associated with Euclidean
$AdS_{3}$. This reformulation facilitates the construction of a new
class of graviton vertex operators, constructed exclusively from entities
arising as worldsheet conformal primaries and Wess-Zumino-Novikov-Witten
(WZNW) currents on the boundary of Euclidean $AdS_{3}$.

This advancement establishes a refined and symmetric correspondence
between celestial CFT and $AdS_{3}$ string theory, wherein gluon
and graviton vertex operators are treated on an equal footing. Their
distinction is thereby confined to the number of supersymmetry generators
and their respective gauge groups, simplifying and unifying the formalism.

The second line of inquiry is devoted to achieving a dynamical realisation
of the framework initially proposed by \citet{de2003holographic}
and subsequently elaborated by \citet{casali2022celestial}. Specifically,
we propose a generalisation of the WZNW model to a supersymmetric
extension of Euclidean $AdS_{3}$, which holographically reproduces
the tree-level MHV scattering amplitudes for gravitons in $\mathcal{N}=8$
Supergravity and for gluons in $\mathcal{N}=4$ supersymmetric Yang-Mills
(SYM) theory.

The derivation of this holographic $AdS_{3}$ theory begins with the
observation that the generating functional for tree-level MHV celestial
amplitudes of gravitons and gluons can be constructed from the chiral
determinant representation of the WZNW action functional. To advance
this formalism, we turn to the framework established by \citet{abe2005multigluon}
and \citet{abe5analysis}. These authors demonstrated that the superspace
constraints characterising the anti-self-dual sector of $\mathcal{N}=8$
Supergravity and $\mathcal{N}=4$ SYM theory can be embedded into
a supersymmetric extension of twistor space, $\mathbf{PT}$. Utilising
ideas from the harmonic superspace formalism developed by \citet{galperin2001harmonic},
it was shown that the superspace constraints in this setting admit
a so-called \emph{chiral semi-analytic gauge}, which simplifies the
constraint equations to those of a WZNW-like field theory. 

Drawing inspiration from these ideas, and employing our observation
that gluon and graviton leaf amplitudes naturally admit an interpretation
on minitwistor space $\mathbf{MT}$, we proceed as follows. We solve
the superspace constraints in the chiral semi-analytic gauge using
the conformal primary basis, and, following \citet{bu2023celestial},
we perform a scaling reduction from twistor space $\mathbf{PT}$ to
minitwistor space $\mathbf{MT}$. This reduction yields explicit forms
for the superpotentials which, when substituted into the WZNW-like
action proposed by \citet{abe2005multigluon}, result in the generating
functionals for graviton and gluon leaf amplitudes.

From this framework, we construct the effective action integral on
$AdS_{3}$ for a WZNW-like field theory, and demonstrate that its
associated Euler-Lagrange equations reproduce the superspace constraints
in the chiral semi-analytic gauge. Moreover, we establish that the
on-shell effective action precisely recovers the generating functionals
for the leaf amplitudes.

\paragraph{Notation. }

We adopt the conventions of \citet{weinberg2010six} concerning the
embedding space formalism, wherein $X^{\mu}$ denotes the Cartesian
coordinates in $\mathbf{R}^{4}$, and $\hat{X}^{\mu}$ represents
the restriction of $X^{\mu}$ to the standard hyperboloid $H_{3}^{+}\subset\mathbf{R}^{4}$.
The coordinates on the $(4\big|\mathcal{N})$-dimensional superspace
$\mathbf{R}^{4|\mathcal{N}}$ are denoted by $\mathbb{X}^{I}\coloneqq(X^{\mu},\theta^{A\alpha},\bar{\theta}_{\alpha}^{\dot{A}})$,
where $\theta^{A\alpha}$ and $\bar{\theta}_{\alpha}^{\dot{A}}$ are
Grassmann-valued, two-component spinors, and $\alpha=1,...,\mathcal{N}$.
Our work here proceeds within the framework of DeWitt supermanifolds
(cf., e.g., \citet{dewitt1992supermanifolds,rogers2007supermanifolds};
for analysis with super-numbers, see \citet{berezin2013introduction}.
For an alternative to DeWitt supermanifolds, refer to \citet{leites1980introduction,batchelor1979structure,batchelor1980two}).

We are concerned, in what follows, with scattering processes involving
gravitons in $\mathcal{N}=8$ Supergravity and gluons in $\mathcal{N}=4$
SYM theory. Lower-case Roman letters $i,j,k...$ will index the $n$
bosons involved in a scattering process. Let $z_{i},\bar{z}_{i}\in\mathbf{CP}^{1}\simeq S^{2}$
denote complex coordinates on the celestial sphere, which can also
be parametrised by pairs of two-component spinors $\pi_{i}^{A}\coloneqq(z_{i},1)^{T}$
and $\bar{\pi}_{i}^{\dot{A}}\coloneqq\left(\bar{z}_{i},1\right)^{T}$,
defining the standard null four-vector $q^{\mu}\left(z_{i},\bar{z}_{i}\right)\coloneqq(\sigma^{\mu})_{A\dot{A}}\pi_{i}^{A}\bar{\pi}_{i}^{\dot{A}}$.
The four-momentum of the $i$-th particle with frequency $s_{i}$
is parametrised by points $z_{i},\bar{z}_{i}\in\mathbf{CP}^{1}$ on
the celestial sphere, using the standard null four-vector as:
\begin{equation}
p^{\mu}=s_{i}q^{\mu}\left(z_{i},\bar{z}_{i}\right)=s_{i}\left(1+z_{i}\bar{z}_{i},z_{i}+\bar{z}_{i},i\left(\bar{z}_{i}-z_{i}\right),1-z_{i}\bar{z}_{i}\right).
\end{equation}
In writing the Berends-Giele-Kuijf (BGK) and Parke-Taylor formulae
for tree-level MHV scattering amplitudes, it is useful to define the
frequency-dependent pairs of two-component spinors $\nu_{i}^{A}\coloneqq\sqrt{s_{i}}\pi_{i}^{A}$
and $\bar{\nu}_{i}^{\dot{A}}\coloneqq\sqrt{s_{i}}\bar{\pi}_{i}^{\dot{A}}$,
such that the four-momentum of the $i$-th graviton takes the form
$p_{i}^{A\dot{A}}=\nu_{i}^{A}\bar{\nu}_{i}^{\dot{A}}$. In celestial
CFT, the conformal weight attributed to the $i$-th graviton is denoted
by $\Delta_{i}$.

\section{$\mathcal{N}=8$ Supergravity \label{sec:Scattering-Amplitudes-in-Minitwistor-Space}}

We begin in Subsection \ref{subsec:Review} by revisiting the construction
of tree-level MHV celestial amplitudes for gravitons as developed
in \citet{mol2024holographic}. Subsequently, in Subsection \ref{subsec:Euclidean--String},
we refine the holographic dictionary proposed in \citet{ogawa2024celestial}
and \citet{mol2024comments}, which establishes a correspondence between
(Euclidean) $AdS_{3}$ string theory and celestial CFT. Finally, in
Subsection \ref{subsec:Generating-Functional}, we derive an expression
for the celestial amplitude for gravitons in $\mathcal{N}=8$ Supergravity,
formulated in terms of multi-graviton wavefunctions on minitwistor
space; this expression serves as our motivation for constructing a
generating functional for graviton amplitudes from the chiral Dirac
determinant.

\subsection{Review\label{subsec:Review}}

Our starting point is the Berends-Giele-Kuijf (BGK) formula for the
tree-level scattering amplitude of $n$-gravitons in the maximally-helicity-violating
configuration $1^{--}$, $2^{--}$, $3^{++}$, ..., $n^{++}$ (for
$n\geq5$) originally introduced by \citet{berends1988relations},
\begin{equation}
\mathcal{M}_{n}=\left(\frac{\kappa}{2}\right)^{n-2}\delta^{\left(4\right)}\left(\sum_{i=1}^{n}p_{i}^{\mu}\right)BGK_{n}+\mathscr{P}_{2,...,n-2},
\end{equation}
where:
\begin{equation}
BGK_{n}=\frac{\left\langle \nu_{1},\nu_{2}\right\rangle ^{8}}{\left\langle \nu_{1},\nu_{2}\right\rangle ...\left\langle \nu_{n},\nu_{1}\right\rangle }\frac{1}{\left\langle \nu_{n},\nu_{1}\right\rangle \left\langle \nu_{1},\nu_{n-1}\right\rangle \left\langle \nu_{n-1},\nu_{n}\right\rangle }\prod_{k=2}^{n-2}\frac{[p_{k}|p_{k+1}+...+p_{n-1}|p_{n}\rangle}{\left\langle \nu_{k},\nu_{n}\right\rangle }.
\end{equation}
Here, $\mathscr{P}_{2,...,n-2}$ denotes the permutation operator
acting on the indices within the set $\{2,...,n-2\}$.

\paragraph{Fermionic Doublet on $\mathbf{CP}^{1}$.}

Following the formalism introduced by \citet{nair1988current,nair2005note},
we incorporate an auxiliary fermionic doublet $(\hat{\chi},\hat{\chi}^{\dagger})$
defined on $\mathbf{CP}^{1}$ (identified as the celestial sphere).
These are introduced through the mode expansions:
\begin{equation}
\hat{\chi}(z_{i})\coloneqq\sum_{k=0}^{\infty}b_{k}z_{i}^{-1-k},\,\,\,\hat{\chi}^{\dagger}(z_{i})\coloneqq\sum_{k=0}^{\infty}b_{k}^{\dagger}z_{i}^{k},
\end{equation}
where the annihilation and creation operators, $b_{k}$ and $b_{k}^{\dagger}$,
respectively, satisfy the anti-commutation relations $\{b_{k},b_{k'}^{\dagger}\}=\delta_{kk'}$
for $k,k'\geq0$, and act on the vacuum state $\big|0\rangle$ as
$b_{k}\big|0\rangle=0$. Consequently, the two-point function of the
doublet $\left(\hat{\chi},\hat{\chi}^{\dagger}\right)$ is given by:
\begin{equation}
\langle0\big|\hat{\chi}(z_{i})\hat{\chi}^{\dagger}(z_{j})\big|0\rangle=\frac{1}{z_{i}-z_{j}}.
\end{equation}
In the subsequent discussion, we introduce the auxiliary two-component
spinor $\lambda^{A}\coloneqq\left(\lambda,1\right)^{T}$. Integration
will be performed over the complex variable $\lambda\in\mathbf{C}$,
restricted to a small contour $C_{n}$ encircling the insertion point
$z_{n}$ of the $n$-th graviton on the celestial sphere $S^{2}\simeq\mathbf{CP}^{1}$. 

We also define an analogous fermionic doublet $(\chi,\chi^{\dagger})$
on the space of two-component spinors. Let $\omega_{i}^{A}\coloneqq(\xi_{i},\zeta_{i})^{T}$
denote a two-component spinor, and consider an open neighbourhood
$\mathcal{U}$ where $\zeta_{i}\neq0$. In this setting, $\omega_{i}^{A}$
can be locally parametrised on $\mathbf{CP}^{1}$ by $z_{i}=\xi_{i}/\zeta_{i}$.
We then define:
\begin{equation}
\chi\left(\omega_{i}\right)\coloneqq\frac{1}{\zeta_{i}}\hat{\chi}\left(z_{i}\right),\,\,\,\chi^{\dagger}(\omega_{i})\coloneqq\frac{1}{\zeta_{i}}\hat{\chi}^{\dagger}(z_{i}).
\end{equation}
This definition ensures that the two-point function of the fermionic
doublet $(\chi,\chi^{\dagger})$ is given by:
\begin{equation}
\langle0\big|\chi(\omega_{i})\chi^{\dagger}(\omega_{j})\big|0\rangle=\frac{1}{\left\langle \omega_{i},\omega_{j}\right\rangle }.
\end{equation}

\paragraph{Nair's Vertex Operators.}

Following the framework elaborated by \citet{nair2005note}, we proceed
to introduce the following operators:
\begin{equation}
\mathcal{Q}_{i}\coloneqq e^{ip_{i}\cdot X}\chi^{\dagger}(\nu_{i})\chi(\nu_{i}),\,\,\,\mathcal{P}_{i}\coloneqq\frac{1}{i}\frac{\bar{\nu}_{i}^{\dot{A}}\lambda^{A}}{\left\langle \nu_{i},\lambda\right\rangle }\frac{\partial}{\partial X^{A\dot{A}}}e^{ip_{i}\cdot X},
\end{equation}
where $\lambda^{A}\coloneqq\left(\lambda,1\right)^{T}$ represents
the auxiliary two-component spinor.

As demonstrated in detail in \citet{mol2024holographic}, these operators
satisfy the identity:
\begin{align}
 & \int\frac{d^{4}X}{\left(2\pi\right)^{4}}\oint_{C_{n}}\frac{d\lambda}{2\pi i}\langle\lambda\big|\mathcal{Q}_{1}\left(\prod_{k=2}^{n-2}\mathcal{P}_{k}\right)\mathcal{Q}_{n-1}\mathcal{Q}_{n}\big|\lambda\rangle\\
 & =-\delta^{\left(4\right)}\left(\sum_{i=1}^{n}p_{i}^{\mu}\right)\frac{1}{\left\langle \nu_{n},\nu_{1}\right\rangle \left\langle \nu_{1},\nu_{n-1}\right\rangle \left\langle \nu_{n-1},\nu_{n}\right\rangle }\prod_{k=2}^{n-2}\frac{[p_{k}|p_{k+1}+...+p_{n-1}|p_{n}\rangle}{\left\langle \nu_{k},\nu_{n}\right\rangle }.
\end{align}
Consequently, the $n$-graviton amplitude can be expressed as:
\begin{align}
\mathcal{M}_{n} & =-\left(\frac{\kappa}{2}\right)^{n-2}\frac{\left\langle \nu_{1},\nu_{2}\right\rangle ^{8}}{\left\langle \nu_{1},\nu_{2}\right\rangle ...\left\langle \nu_{n},\nu_{1}\right\rangle }\int\frac{d^{4}X}{\left(2\pi\right)^{4}}\oint_{C_{n}}\frac{d\lambda}{2\pi i}\langle\lambda\big|\mathcal{Q}_{1}\left(\prod_{k=2}^{n-2}\mathcal{P}_{k}\right)\mathcal{Q}_{n-1}\mathcal{Q}_{n}\big|\lambda\rangle\label{eq:Step-14}\\
 & +\mathscr{P}_{2,...,n-2}.
\end{align}

\paragraph{Celestial Amplitudes.}

The present formalism now departs from the construction proposed by
\citet{nair2005note}, as we now prepare to perform the Mellin transform
and derive the celestial graviton amplitude. Reformulating Eq. (\ref{eq:Step-14})
in terms of the frequencies $s_{i}$ and two-component spinors $\pi_{i}^{A}\coloneqq\left(z_{i},1\right)^{T}$,
the amplitude takes the form:
\begin{align}
\mathcal{M}_{n} & =-\left(\frac{\kappa}{2}\right)^{n-2}\prod_{i=1}^{n}s_{i}^{e_{i}}\frac{\left\langle \pi_{1},\pi_{2}\right\rangle ^{8}}{\left\langle \pi_{1},\pi_{2}\right\rangle ...\left\langle \pi_{n},\pi_{1}\right\rangle }\int\frac{d^{4}X}{\left(2\pi\right)^{4}}\oint_{C_{n}}\frac{d\lambda}{2\pi i}\langle\lambda\big|\mathcal{Q}_{1}\left(\prod_{k=2}^{n-2}\mathcal{P}_{k}\right)\mathcal{Q}_{n-1}\mathcal{Q}_{n}\big|\lambda\rangle\\
 & +\mathscr{P}_{2,...,n-2},
\end{align}
where the exponents for the chosen MHV configuration are given by
$e_{1}=e_{2}=3$ and $e_{3}=...=e_{n}=-1$. Noting that:
\begin{equation}
\mathcal{Q}_{i}=\frac{1}{s_{i}}e^{is_{i}q\left(z_{i},\bar{z}_{i}\right)\cdot X}\hat{\chi}^{\dagger}(z_{i})\hat{\chi}\left(z_{i}\right),\,\,\,\mathcal{P}_{i}=\frac{1}{i}\frac{\bar{\pi}_{i}^{\dot{A}}\lambda^{A}}{\left\langle \pi_{i},\lambda\right\rangle }\frac{\partial}{\partial X^{A\dot{A}}}e^{is_{i}q\left(z_{i},\bar{z}_{i}\right)\cdot X}.
\end{equation}
the Mellin-transformed operators are given by:
\begin{equation}
\widehat{\mathcal{Q}}_{i}\coloneqq\int_{\left(0,\infty\right)}\frac{ds_{i}}{s_{i}}s_{i}^{\Delta_{i}}\left(s_{i}^{e_{i}}\mathcal{Q}_{i}\right)=\phi_{2h_{i}}\left(z_{i},\bar{z}_{i}\big|X\right)\hat{\chi}^{\dagger}(z_{i})\hat{\chi}\left(z_{i}\right),
\end{equation}
\begin{equation}
\widehat{\mathcal{P}}_{i}\coloneqq\int_{\left(0,\infty\right)}\frac{ds_{i}}{s_{i}}s_{i}^{\Delta_{i}}\left(s_{i}\mathcal{P}_{i}\right)=\frac{1}{i}\frac{\bar{\pi}_{i}^{\dot{A}}\lambda^{A}}{\left\langle \pi_{i},\lambda\right\rangle }\frac{\partial}{\partial X^{A\dot{A}}}\phi_{2h_{i}}\left(z_{i},\bar{z}_{i}\big|X\right).
\end{equation}
Here, following \citet{pasterski2017conformal}, we introduced the
celestial conformal primary wavefunction for massless scalars:
\begin{equation}
\phi_{2h_{i}}\left(z_{i},\bar{z}_{i}\big|X\right)\coloneqq\frac{\Gamma\left(2h_{i}\right)}{\left(\varepsilon-iq\left(z_{i},\bar{z}_{i}\right)\cdot X\right)},
\end{equation}
with $\varepsilon>0$ serving as a regulator. For the MHV configuration
$1^{--}$, $2^{--}$, $3^{++}$, ..., $n^{++}$, the scaling dimensions
are defined as follows: \label{Scaling-Dimensions}
\begin{equation}
2h_{i}\coloneqq\Delta_{i}+e_{i}-1,\,\,\,i\in\{1,n-1,n\};\,\,2h_{i}\coloneqq\Delta_{i}+e_{i},\,\,\,i\in\{2,...,n-2\}.\label{eq:Definitions}
\end{equation}

The celestial $n$-amplitude is the $\varepsilon$-regulated Mellin
transform of $\mathcal{M}_{n}$:
\begin{equation}
\widehat{\mathcal{M}}_{n}=\prod_{k=1}^{n}\int_{\left(0,\infty\right)}\frac{ds_{k}}{s_{k}}s_{k}^{\Delta_{k}}e^{-\varepsilon s_{k}}\mathcal{M}_{n}.
\end{equation}
Therefore:
\begin{align}
\widehat{\mathcal{M}}_{n} & =-\left(\frac{\kappa}{2}\right)^{n-2}\frac{\left\langle \pi_{1},\pi_{2}\right\rangle ^{8}}{\left\langle \pi_{1},\pi_{2}\right\rangle ...\left\langle \pi_{n},\pi_{1}\right\rangle }\int\frac{d^{4}X}{\left(2\pi\right)^{4}}\oint_{C_{n}}\frac{d\lambda}{2\pi i}\langle\lambda\big|\widehat{\mathcal{Q}}_{1}\left(\prod_{k=2}^{n-2}\widehat{\mathcal{P}}_{k}\right)\widehat{\mathcal{Q}}_{n-1}\widehat{\mathcal{Q}}_{n}\big|\lambda\rangle\label{eq:Step-15}\\
 & +\mathscr{P}_{2,...,n-2}.
\end{align}

\paragraph{Supersymmetry.}

The transition from General Relativity to $\mathcal{N}=8$ Supergravity
begins with the following observation. Let $\theta^{A\alpha}$ $\left(\alpha=1,...,8\right)$
denote Grassmann-valued two-component spinors normalised such that:
\begin{equation}
\int d^{2}\theta\theta_{A}^{\alpha}\theta_{B}^{\alpha}=\varepsilon_{AB},
\end{equation}
where the Berezin integral is defined as per \citet{berezin2013introduction}
and \citet{dewitt1992supermanifolds}. By defining the coefficients:
\begin{equation}
\xi_{i}^{\alpha}\coloneqq\theta^{A\alpha}\pi_{iA}\,\,\,\left(1\leq i\leq n\right),\label{eq:Definition}
\end{equation}
one obtains:
\begin{equation}
\int d^{16}\theta\prod_{\alpha=1}^{8}\xi_{i}^{\alpha}\xi_{j}^{\alpha}=\left\langle \pi_{i},\pi_{j}\right\rangle ^{8}.
\end{equation}
Then, Eq. (\ref{eq:Step-15}) can be rewritten as:
\begin{align}
\widehat{\mathcal{M}}_{n} & =-\left(\frac{\kappa}{2}\right)^{n-2}\int d^{16}\theta\left(\prod_{\alpha=1}^{8}\xi_{1}^{\alpha}\xi_{2}^{\alpha}\right)\prod_{i=1}^{n}\frac{1}{\left\langle \pi_{i},\pi_{i+1}\right\rangle }\label{eq:Step-15-1}\\
 & \int\frac{d^{4}X}{\left(2\pi\right)^{4}}\oint_{C_{n}}\frac{d\lambda}{2\pi i}\langle\lambda\big|\widehat{\mathcal{Q}}_{1}\left(\prod_{k=2}^{n-2}\widehat{\mathcal{P}}_{k}\right)\widehat{\mathcal{Q}}_{n-1}\widehat{\mathcal{Q}}_{n}\big|\lambda\rangle+\mathscr{P}_{2,...,n-2}.
\end{align}
Here, we set $\pi_{n+1}\coloneqq\pi_{1}$, so that $\prod_{k=1}^{n}\left\langle \pi_{k},\pi_{k+1}\right\rangle =z_{12}z_{23}...z_{n1}$.

This naturally motivates the introduction of the $\mathcal{N}=8$
superfield:\label{Superfield}
\begin{equation}
\Psi_{i}\coloneqq h^{-}(z_{i},\bar{z}_{i})+\xi^{\alpha}\tilde{\lambda}_{\alpha}+\frac{1}{2}\xi^{\alpha}\xi^{\beta}\widetilde{\varphi}_{\alpha\beta}+...+h^{+}(z_{i},\bar{z}_{i})\prod_{\alpha=1}^{8}\xi_{i}^{\alpha},\label{eq:Superfield}
\end{equation}
which contains the degrees of freedom $(h^{\pm},\tilde{\lambda}_{\alpha},\widetilde{\phi}_{\alpha\beta},...)$
of $\mathcal{N}=8$ Supergravity. Here, $h_{i}^{-}\coloneqq h^{-}\left(z_{i},\bar{z}_{i}\right)$
and $h_{i}^{+}\coloneqq h^{+}\left(z_{i},\bar{z}_{i}\right)$ correspond
to the vacuum expectation values of the annihilation operators for
gravitons of negative and positive helicities, respectively. The dependence
of $\Psi_{i}$ on the holomorphic and anti-holomorphic coordinates
$z_{i},\bar{z}_{i}$ is encoded through the coefficients $\xi_{i}^{\alpha}$
$\left(\alpha=1,...,8\right)$ defined in Eq. (\ref{eq:Definition}). 

Therefore, we pass from the celestial amplitude $\widehat{\mathcal{M}}_{n}$
to the \emph{celestial partial} \emph{super-amplitude} $\widehat{\mathbb{M}}_{n}$:
\begin{align}
\widehat{\mathbb{M}}_{n} & =-\frac{1}{\left(2\pi\right)^{4}}\left(\frac{\kappa}{2}\right)^{n-2}\int d^{4}X\int d^{16}\theta\oint_{C_{n}}\frac{d\lambda}{2\pi i}\langle\lambda\big|\widehat{\mathcal{Q}}_{1}\left(\prod_{k=2}^{n-2}\widehat{\mathcal{P}}_{k}\right)\widehat{\mathcal{Q}}_{n-1}\widehat{\mathcal{Q}}_{n}\big|\lambda\rangle\prod_{\ell=1}^{n}\frac{\Psi_{\ell}}{\left\langle \pi_{\ell},\pi_{\ell+1}\right\rangle }\\
 & +\mathscr{P}_{2,...,n-2}.
\end{align}

\paragraph{Factorisation of Celestial Wavefunctions.}

To incorporate the formalism of leaf amplitudes as developed by \citet{casali2022celestial}
and \citet{melton2023celestial}, we first analytically continue $\widehat{\mathbb{M}}_{n}$
from Lorentzian to Kleinian signature via the transformation $X^{2}\mapsto-iX^{2}$.
Then, the celestial amplitude must be expressed as a weighted integral
over the leaves of the natural hyperbolic foliation of Klein space
$\mathbf{R}^{\left(2,2\right)}$, in terms of the Kleinian hyperboloid
$\mathbf{H}_{3}\coloneqq AdS_{3}/\mathbf{Z}$. Achieving this requires
factoring the celestial wavefunctions $\phi_{2h_{i}}\left(z_{i},\bar{z}_{i}\big|X\right)$
contained within the operators $\widehat{\mathcal{Q}}_{i}$ and $\widehat{\mathcal{P}}_{i}$
outside the correlator $\langle\lambda\big|...\big|\lambda\rangle$.

To facilitate this, we define the following pair of operators:
\begin{equation}
\mathsf{A}_{i}\coloneqq\exp\left(q\left(z_{i},\bar{z}_{i}\right)\cdot y\,\mathsf{P}_{i}\right)\hat{\chi}^{\dagger}(z_{i})\hat{\chi}\left(z_{i}\right),\,\,\,\mathsf{B}_{i}\coloneqq\frac{\bar{\pi}_{i}^{\dot{A}}\lambda^{A}}{\left\langle \pi_{i},\lambda\right\rangle }\frac{\partial}{\partial y^{A\dot{A}}}\exp\left(q\left(z_{i},\bar{z}_{i}\right)\cdot y\,\mathsf{P}_{i}\right),\label{eq:Definitions-1}
\end{equation}
where $\mathsf{P}_{i}$ is a weight-shifting operator that acts on
celestial wavefunctions by:
\begin{equation}
\mathsf{P}_{i}\phi_{2h_{j}}\left(z_{j},\bar{z}_{j}\big|X\right)\coloneqq\phi_{2h_{j}+\delta_{ij}}\left(z_{j},\bar{z}_{j}\big|X\right),
\end{equation}
and $y^{A\dot{A}}$ is a four-vector that parametrises the family
of operators $\mathsf{A}_{i},\mathsf{B}_{i}$.

Consider now the following identity:
\begin{align}
 & \frac{1}{i}\frac{\bar{\pi}_{i}^{\dot{A}}\lambda^{A}}{\left\langle \pi_{i},\lambda\right\rangle }\frac{\partial}{\partial X^{A\dot{A}}}\phi_{2h_{j}}\left(z_{j},\bar{z}_{j}\big|X\right)\\
 & =\int d^{4}y\delta^{\left(4\right)}\left(y\right)\frac{\bar{\pi}_{i}^{\dot{A}}\lambda^{A}}{\left\langle \pi_{i},\lambda\right\rangle }\frac{\partial}{\partial y^{A\dot{A}}}\exp\left(q\left(z_{j},\bar{z}_{j}\right)\cdot y\,\mathsf{P}_{j}\right)\phi_{2h_{j}}\left(z_{j},\bar{z}_{j}\big|X\right).
\end{align}
For the sake of notational economy, henceforth we shall omit the integral
$\int d^{4}y\delta^{\left(4\right)}\left(y\right)$. Accordingly,
all subsequent equalities involving the operators $\mathsf{A}_{i}$
and $\mathsf{B}_{i}$ should be understood as being evaluated at $y^{A\dot{A}}=0$.

Under this convention, the celestial super-amplitude can be reformulated
in terms of these operators as follows:
\begin{align}
\widehat{\mathbb{M}}_{n} & =-\frac{1}{\left(2\pi\right)^{4}}\left(\frac{\kappa}{2}\right)^{n-2}\oint_{C_{n}}\frac{d\lambda}{2\pi i}\langle\lambda\big|\mathsf{A}_{1}\left(\prod_{k=2}^{n-2}\mathsf{B}_{k}\right)\mathsf{A}_{n-1}\mathsf{A}_{n}\big|\lambda\rangle\label{eq:Step-16}\\
 & \int d^{4}X\int d^{16}\theta\prod_{\ell=1}^{n}\frac{\phi_{2h_{\ell}}\left(z_{\ell},\bar{z}_{\ell}\big|X\right)\Psi_{\ell}}{\left\langle \pi_{\ell},\pi_{\ell+1}\right\rangle }+\mathscr{P}_{2,...,n-2}.
\end{align}
As desired, \emph{this representation explicitly factors out the product
$\prod_{\ell}\phi_{2h_{\ell}}(z_{\ell},\bar{z}_{\ell}|x)$ of celestial
wavefunctions from the correlator} $\langle\lambda\big|\cdot\cdot\cdot\big|\lambda\rangle$. 

\subsection{Euclidean $AdS_{3}$ String Theory and Graviton Leaf Amplitudes\label{subsec:Euclidean--String}}

We now proceed to derive a novel expression for the tree-level MHV
celestial super-amplitude for gravitons within $\mathcal{N}=8$ Supergravity.
This formula will facilitate the construction of an improved correspondence
between (Euclidean) $AdS_{3}$ string theory and celestial CFT, as
originally suggested by \citet{ogawa2024celestial} and \citet{mol2024comments}.

\paragraph{Celestial Leaf Amplitudes.}

Utilising the formalism of celestial leaf amplitudes as developed
by \citet{melton2023celestial}, \citet{melton2024celestial} and
reviewed in the Appendix of \citet{mol2024holographic}, we examine
the spacetime integral, $\int d^{4}X\,(\cdot)$, appearing in Eq.
(\ref{eq:Step-16}). Under the simplifying assumption that all gravitons
are outgoing, this integral admits the following representation:
\begin{equation}
\int d^{4}X\prod_{k=1}^{n}\phi_{2h_{k}}\left(z_{k},\bar{z}_{k}\big|X\right)=2\pi\delta\left(\beta\right)\int_{\mathbf{H}_{3}}d^{3}x\prod_{k=1}^{n}G_{2h_{k}}(z_{k},\bar{z}_{k}|x)+\left(\bar{z}_{k}\rightarrow-\bar{z}_{k}\right)\label{eq:Step-11}
\end{equation}
where the function $x\in\mathbf{H}_{3}\mapsto G_{\Delta}\left(z,\bar{z}\big|x\right)$
is given by:
\begin{equation}
G_{\Delta}\left(z,\bar{z}\big|x\right)\coloneqq\frac{\Gamma\left(\Delta\right)}{\left(\varepsilon-iq\left(z,\bar{z}\right)\cdot x\right)^{\Delta}},
\end{equation}
and corresponds to the bulk-to-boundary Green's function for the covariant
Laplacian on the hyperboloid $\mathbf{H}_{3}$ (cf. \citet[Appendix A]{teschner1999mini},
\citet{penedones2017tasi} and \citet{costa2014spinning}). The symbol
$\left(\bar{z}_{k}\rightarrow-\bar{z}_{k}\right)$ indicates the repetition
of the first term under the replacement $\bar{z}_{k}\mapsto-\bar{z}_{k}$
for all $1\leq k\leq n$. The parameter $\beta$, defined by:
\begin{equation}
\beta\coloneqq2\sum_{i=1}^{n}h_{i}-4,
\end{equation}
encodes the total scaling dimension associated with the scattering
process under consideration. Accordingly, the integral $\int d^{3}x\,(\cdot)$
over the hyperboloid $\mathbf{H}_{3}$ appearing on the right-hand
side of Eq. (\ref{eq:Step-11}) can be interpreted physically as a
contact Feynman-Witten diagram representing the propagation of massless
scalar particles in Euclidean $AdS_{3}$.

\paragraph{The Weight-shifting Polynomial Operator.}

To simplify subsequent computations, we define the following operator,
acting on the set $\{\phi_{2h_{i}}(z_{i},\bar{z}_{i}\big|X)\}$ of
celestial conformal primaries:
\begin{equation}
\fint dh_{1}...dh_{n}\coloneqq\frac{i}{2\pi}\oint_{C_{n}}d\lambda\langle\lambda\big|\mathsf{A}_{1}\left(\prod_{k=2}^{n-2}\mathsf{B}_{k}\right)\mathsf{A}_{n-1}\mathsf{A}_{n}\big|\lambda\rangle.
\end{equation}
Recalling the definitions of the operators $\mathsf{A}_{i}$ and $\mathsf{B}_{i}$,
as introduced in Eq. (\ref{eq:Definitions-1}), and employing an inductive
argument, \emph{the operator $\fint dh_{1}...dh_{n}$ can be expanded
as a polynomial in the weight-shifting operators} $\mathsf{P}_{i}$:
\begin{equation}
\fint dh_{1}...dh_{n}=\frac{1}{\left(z_{n}-z_{1}\right)\left(z_{1}-z_{n-1}\right)\left(z_{n-1}-z_{n}\right)}\left(\prod_{i=2}^{n-2}\sum_{j_{i}=i+1}^{n}\right)\prod_{k=2}^{n-2}\frac{\left(\bar{z}_{k}-\bar{z}_{j_{k}}\right)\left(z_{j_{k}}-z_{n}\right)}{z_{k}-z_{n}}\mathsf{P}_{j_{k}}.
\end{equation}
Since $\mathsf{P}_{1},...,\mathsf{P}_{n}$ acts by shifting the scaling
dimensions $2h_{1},...,2h_{n}$, this expression provides a justification
for the mnemonic symbol $\fint dh_{1}...dh_{n}$ employed in its definition.

Finally, the celestial super-amplitude $\widehat{\mathbb{M}}_{n}$
can be elegantly rewritten as:
\begin{equation}
\widehat{\mathbb{M}}_{n}=\frac{1}{\left(2\pi\right)^{3}}\left(\frac{\kappa}{2}\right)^{n-2}\fint dh_{1}...dh_{n}\delta\left(\beta\right)M_{n}+\left(\bar{z}_{k}\rightarrow-\bar{z}_{k}\right)+\mathscr{P}_{2,...,n-2}
\end{equation}
where, following \citet{melton2023celestial}, the leaf amplitude
$M_{n}$ is defined by:
\begin{equation}
M_{n}\coloneqq\int_{\mathbf{H}_{3}}d^{3}x\int d^{16}\theta\prod_{k=1}^{n}G_{2h_{k}}\left(z_{k},\bar{z}_{k}\big|x\right)\frac{\Psi_{k}}{\left\langle \pi_{k},\pi_{k+1}\right\rangle }.
\end{equation}

\begin{rem*}
The foregoing expression describes the tree-level MHV leaf amplitude
for gravitons within $\mathcal{N}=8$ Supergravity, and notably, it
relies exclusively on the kinematical data parametrised by points
$x\in\mathbf{H}_{3}$ on the hyperboloid, and $\left(z_{i},\bar{z}_{i}\right)\in\mathbf{CP}^{1}$,
representing coordinates on the celestial sphere. This construction
thus achieves the sought-after reformulation of the celestial super-amplitude
in the desired geometric and kinematic variables. 
\end{rem*}
\begin{rem*}
The graviton leaf amplitude $M_{n}$ can be equivalently expressed
in the form:
\begin{equation}
M_{n}=\prod_{i=1}^{n}\int d^{2}\mu_{i}\frac{\Gamma\left(2h_{i}\right)}{\left(-i\mu_{i}\cdot\bar{\pi}_{i}\right)}\int_{\mathbf{H}_{3}}d^{3}x\int d^{16}\theta\,\psi,
\end{equation}
where:
\begin{equation}
\psi\coloneqq\prod_{i=1}^{n}\delta^{\left(2\right)}\left(\mu_{\dot{A}i}-\pi_{i}^{A}x_{A\dot{A}}\right)\frac{\Psi_{i}}{\left\langle \pi_{i},\pi_{i+1}\right\rangle }.
\end{equation}
Following the terminology introduced by \citet{cachazo2013gravity,cachazo2014gravity}
and \citet{adamo2013twistor}, the function $\psi$ may be understood
as the \emph{multi-graviton wavefunction} constructed on the minitwistor
space $(\mu_{\dot{A}i},\pi_{i}^{A})\in\mathbf{MT}$ associated with
the hyperboloid $\mathbf{H}_{3}$. From this perspective, the geometry
of minitwistor space provides a framework for encoding the kinematic
data of \emph{leaf} amplitudes. For a review of the minitwistor space
$\mathbf{MT}$ in the context of celestial holography, cf. \citet{bu2023celestial}. 
\end{rem*}

\paragraph{Euclidean $AdS_{3}$ String Theory.}

The connection between celestial CFT and (Euclidean) $AdS_{3}$ string
theory can be introduced through the following observations. Let $\Phi^{h_{i}}\left(w_{i}\big|z_{i}\right)$
denote the conformal primary fields of the level-$k$ $H_{3}^{+}$-WZNW
model, characterised by spin $h_{i}$, where $w_{i},\bar{w}_{i}\in\Sigma$
are worldsheet coordinates, and $z_{i},\bar{z}_{i}\in\mathbf{CP}^{1}$
are coordinates on the celestial sphere. For notational convenience,
we suppress the dependence of the conformal primaries on the anti-holomorphic
coordinates $\bar{w}_{i}$ and $\bar{z}_{i}$.

It was demonstrated by \citet{teschner1999mini} and subsequently
refined by \citet{ribault2005h3+} that, in the minisuperspace limit,
defined by $k\rightarrow\infty$, the correlation functions of the
conformal primaries in the $H_{3}^{+}$-WZNW model can be expressed
in terms of the bulk-to-boundary Green's functions on $\mathbf{H}_{3}$
as follows:
\begin{equation}
\lim_{k\rightarrow\infty}\prod_{i=1}^{n}\int d^{2}w_{i}\Gamma\left(2h_{i}\right)\langle\Phi^{h_{1}}\left(w_{1}\big|z_{1}\right)...\Phi^{h_{n}}\left(w_{n}\big|z_{n}\right)\rangle=\int_{\mathbf{H}_{3}}d^{3}x\prod_{i=1}^{n}G_{2h_{i}}\left(z_{i},\bar{z}_{i}\big|x\right).
\end{equation}
To elaborate: the minisuperspace limit of the correlation functions
of the primary fields in the $H_{3}^{+}$-WZNW model yields the computation
of a contact Feynman-Witten diagram on the three-dimensional hyperboloid.

We now consider (Euclidean) $AdS_{3}$ string theory formulated on
the target space $AdS_{3}\times X$, where $X$ is a compact manifold.
Let $J^{a}\left(z\right)$ be a level-one $SO\left(2N\right)$ WZNW
current algebra defined on the boundary of $AdS_{3}$. It is postulated
that $J^{a}\left(z\right)$ originates from the CFT on $X$, henceforth
referred to as the $X$-CFT.

Let $\mathsf{T}^{a}$ represent the generators of the adjoint representation
of $SO\left(2N\right)$, normalised such that $\text{Tr}\left(\mathsf{T}^{a}\mathsf{T}^{b}\right)=2\delta^{ab}$,
with the commutation relations $\left[\mathsf{T}^{a},\mathsf{T}^{b}\right]=if^{abc}\mathsf{T}^{c}$.
At leading-trace order, the correlators of the WZNW currents $J^{a}\left(z\right)$
satisfy the following identity:
\begin{equation}
\langle J^{a_{1}}(z_{1})...J^{a_{n}}\left(z_{n}\right)\rangle\sim\frac{\text{Tr}\left(\mathsf{T}^{a_{1}}...\mathsf{T}^{a_{n}}\right)}{z_{12}z_{23}...z_{n1}}=\text{Tr}\prod_{i=1}^{n}\frac{\mathsf{T}^{a_{i}}}{\left\langle \pi_{i},\pi_{i+1}\right\rangle }.
\end{equation}
Hereafter, all equations involving the correlation functions of the
WZNW currents $J^{a}\left(z\right)$ are to be understood at the leading-trace
order.

Subsequently, we define the currents $\mathcal{J}\left(z\right)\coloneqq\mathsf{T}^{a}J^{a}\left(z\right)$,
such that:
\begin{equation}
\langle\mathcal{J}\left(z_{1}\right)...\mathcal{J}\left(z_{n}\right)\rangle=\mathcal{N}\prod_{i=1}^{n}\frac{1}{\left\langle \pi_{i},\pi_{i+1}\right\rangle }.
\end{equation}
The proportionality factor is given by $\mathcal{N}\coloneqq\text{Tr}\left(\mathsf{T}^{a_{1}}...\mathsf{T}^{a_{n}}\right)^{2}$.
Building on the foundational works of \citet{giveon1998comments,dolan1999vertex,dolan2001vertex},
let $W_{X}$ denote a spinless operator in the $X$-CFT. One can construct
physical vertex operators of the form $W_{X}V_{jm\bar{m}}$, where
$V_{jm\bar{m}}$ represents a worldsheet vertex operator. 

In particular, we define the \emph{celestial vertex operators for
gravitons} as:
\begin{equation}
\mathcal{V}_{i}\coloneqq\int d^{2}w_{i}\Gamma\left(2h_{i}\right)\mathcal{J}\left(z_{i}\right)\Psi_{i}\Phi^{h_{i}}\left(w_{i}\big|z_{i}\right).
\end{equation}

Finally, the \emph{leaf amplitude} $M_{n}$ for graviton scattering
in $\mathcal{N}=8$ Supergravity can then be expressed as the minisuperspace
limit $\left(k\rightarrow\infty\right)$ of (Euclidean) $AdS_{3}$
string theory as:
\begin{equation}
M_{n}=\frac{1}{\mathcal{N}}\lim_{k\rightarrow\infty}\int d^{16}\theta\left\langle \mathcal{V}_{1}...\mathcal{V}_{n}\right\rangle .
\end{equation}
This result provides a refined formulation of the correspondence between
the sector of celestial CFT encoding the tree-level MHV graviton scattering
amplitudes and (Euclidean) $AdS_{3}$ string theory. Notably, in contrast
to our earlier proposal (\citet{mol2024comments}), the celestial
vertex operators $\mathcal{V}_{i}$ for gravitons are entirely constructed
from objects derived as solutions to the string theory equations of
motion, either worldsheet conformal primaries or currents from the
$X$-CFT. Crucially, it is no longer necessary to ``dress'' the operators
$\mathcal{V}_{i}$ with any additional \emph{ad hoc} components.

\subsection{Generating Functional\label{subsec:Generating-Functional}}

We now undertake the derivation of a generating functional for the
graviton leaf amplitude $M_{n}$ using the WZNW action. The result
herein constitutes a key step in our construction of a holographic
$AdS_{3}$ field theory dual to the sector of celestial CFT containing
$\mathcal{N}=8$ Supergravity. The derivation rests upon two basic
observations.

Firstly, let $\mathcal{S}[G]$ denote the action integral of the level-one
$SO\left(2N\right)$ WZNW model defined on $\mathbf{CP}^{1}$. The
dynamical variable $G=G\left(z,\bar{z}\right)$ is a $\mathbf{CP}^{1}$-dependent
matrix-valued field in the Lie group $SO\left(2N\right)$. From $G$,
we construct a differential one-form $\boldsymbol{\omega}$ on $\mathbf{CP}^{1}$,
\begin{equation}
\boldsymbol{\omega}\coloneqq\omega_{z}dz+\omega_{\bar{z}}d\bar{z},\label{eq:Differential}
\end{equation}
where the holomorphic and anti-holomorphic components are defined
as follows:
\begin{equation}
\omega_{z}\coloneqq-\left(\partial_{z}G\right)G^{-1},\,\,\,\omega_{\bar{z}}\coloneqq-\left(\partial_{\bar{z}}G\right)G^{-1}.
\end{equation}
As reviewed by \citet{nair2005chern}, if $G$ satisfies the Euler-Lagrange
equations associated with the WZNW action $\mathcal{S}[G]$, then
the one-form $\boldsymbol{\omega}$ obeys the condition:
\begin{equation}
\mathcal{F}[\boldsymbol{\omega}]\coloneqq\partial_{z}\omega_{\bar{z}}-\partial_{\bar{z}}\omega_{z}+[\omega_{z},\omega_{\bar{z}}]=0.
\end{equation}
This equation suggests that $\boldsymbol{\omega}$ can be interpreted
as the gauge potential corresponding to a connection one-form on $\mathbf{CP}^{1}$,
valued in the Lie algebra $\mathfrak{so}\left(2N\right)$. The term
$\mathcal{F}[\boldsymbol{\omega}]$ then represents the curvature
two-form, which quantifies the field strength of the gauge potential
$\boldsymbol{\omega}$. Consequently, the vanishing of the field strength,
$\mathcal{F}[\boldsymbol{\omega}]=0$, can be identified with the
``flatness'' condition for the connection defined by $\boldsymbol{\omega}$.

The second observation proceeds as follows. Let $\mathcal{D}_{z}\coloneqq\partial_{z}+[\omega_{z},\cdot]$
denote the holomorphic component of the gauge-covariant derivative
operator induced by the connection $\boldsymbol{\omega}$. As explained
by \citet{nair2005chern}, the WZNW action $\mathcal{S}[G]$ admits
an alternative representation as a chiral determinant:
\begin{equation}
\mathcal{S}[G]=\mathcal{S}[\boldsymbol{\omega}]\coloneqq\text{Tr}\log\mathcal{D}_{z}-\text{Tr}\log\omega_{z}.
\end{equation}
This expression can be formally expanded as:
\begin{equation}
\mathcal{S}[\boldsymbol{\omega}]=\sum_{m\geq2}\frac{\left(-1\right)^{m+1}}{m}\int\frac{d^{2}z_{1}}{\pi}...\int\frac{d^{2}z_{m}}{\pi}\text{Tr}\left(\frac{\omega_{z}\left(z_{1},\bar{z}_{1}\right)...\omega_{z}\left(z_{m},\bar{z}_{m}\right)}{z_{12}z_{23}...z_{m1}}\right).\label{eq:Formal-Expansion}
\end{equation}

Now, consider the possibility of lifting the connection one-form $\boldsymbol{\omega}$
from $\mathbf{CP}^{1}$ to a holomorphic extension over the $\mathcal{N}=8$
supersymmetric hyperboloid $\mathbf{H}_{3}$. Denote this lifted connection
by $\boldsymbol{\Omega}\coloneqq\Omega_{z}dz+\Omega_{\bar{z}}dz$,
with the holomorphic component defined as: 
\begin{equation}
\Omega_{z}(z_{i},\bar{z}_{i}\big|x,\theta,\bar{\theta})\coloneqq\int_{\mathcal{P}}\frac{d\Delta_{i}}{2\pi i}G_{2h_{i}}\left(z_{i},\bar{z}_{i}\big|x\right)\Psi_{i}.\label{eq:Postulate}
\end{equation}
Here, $\mathcal{P}\coloneqq1+i\mathbf{R}$, and the scaling dimension
$2h_{i}$ depends affinely on $\Delta_{i}$, as specified in Eq. (\ref{eq:Definitions}).
Substituting this expression into the expanded form of $\mathcal{S}[\boldsymbol{\omega}]$,
we obtain:
\begin{equation}
\mathcal{S}[\boldsymbol{\Omega}]=\sum_{m\geq2}\frac{\left(-1\right)^{m+1}}{m}\left(\prod_{k=1}^{n}\int_{\mathcal{P}}\frac{d\Delta_{k}}{2\pi i}\int\frac{d^{2}z_{k}}{\pi}\right)\prod_{i=1}^{n}\frac{\Omega_{z}(z_{i},\bar{z}_{i}\big|x,\theta,\bar{\theta})}{\left\langle \pi_{i},\pi_{i+1}\right\rangle }.
\end{equation}

We then define the ``effective'' action functional by:
\begin{equation}
\exp(\mathcal{W}[\boldsymbol{\Omega}])\coloneqq\int_{\mathbf{H}_{3}}d^{3}x\int d^{16}\theta\,\mathcal{S}[\boldsymbol{\Omega}].
\end{equation}
Taking the functional derivatives of $e^{\mathcal{W}[\boldsymbol{\Omega}]}$
with respect to $h_{i}^{\pm}$, the vacuum expectation values of annihilation
operators for gravitons with positive and negative helicities (contained
in the $\mathcal{N}=8$ superfield $\Psi_{i}$, as defined in Eq.
(\ref{eq:Superfield}), Subsection \ref{Superfield}) yields:
\begin{equation}
M_{n}=\frac{\delta}{\delta h_{1}^{-}}\frac{\delta}{\delta h_{2}^{-}}\frac{\delta}{\delta h_{3}^{+}}...\frac{\delta}{\delta h_{n}^{+}}\exp(\mathcal{W}[\boldsymbol{\Omega}])\Big|_{\boldsymbol{\Omega}=0}.\label{eq:Generating-Functional}
\end{equation}
Thus, the effective action $\mathcal{W}[\boldsymbol{\Omega}]$ serves
as the generating functional for the graviton leaf amplitude $M_{n}$
in $\mathcal{N}=8$ Supergravity.

Consequently, the construction of the holographic $AdS_{3}$ theory
proceeds through the following steps:
\begin{enumerate}
\item \emph{Postulate an Action Integral.}\textbf{ }Introduce an action
$\mathcal{I}[G,\boldsymbol{\Omega}]$ such that its variation with
respect to the matrix field $G$ imposes the ``flatness'' condition
$\mathcal{F}[\boldsymbol{\Omega}]=0$, and its on-shell effective
action coincides with $\mathcal{W}[\boldsymbol{\Omega}]$.
\item \emph{Lift the Connection.}\textbf{ }Establish a mathematical procedure
to justify the lifting of the $\mathbf{CP}^{1}$ connection $\boldsymbol{\omega}$
to the holomorphically extended $\mathcal{N}=8$ supersymmetric hyperboloid
$\mathbf{H}_{3}$.
\end{enumerate}
The procedure for lifting the $\mathbf{CP}^{1}$ connection $\boldsymbol{\omega}$
to its extension $\boldsymbol{\Omega}$ on $\mathbf{H}_{3}$ is predicated
upon an important insight by \citet{abe2005multigluon} and \citet{abe5analysis}.
Drawing inspiration from the harmonic superspace formalism, comprehensively
reviewed in \citet{galperin2001harmonic}, these authors demonstrated
that the superspace constraints of $\mathcal{N}=4$ supersymmetric
Yang-Mills theory and the anti-self-dual sector of $\mathcal{N}=8$
Supergravity can be embedded into a supersymmetric generalisation
of twistor space $\mathbf{PT}$. By imposing the so-called \emph{chiral
semi-analytic gauge}, these constraints are shown to reduce exactly
to the ``flatness'' condition $\mathcal{F}[\boldsymbol{\Omega}]=0$
introduced in the preceding discussion. Building upon their work,
we shall prove that a scaling-reduction of the solution to the \emph{analyticity
constraint} results precisely in the lifting $\boldsymbol{\Omega}$
of the $\mathbf{CP}^{1}$ connection $\boldsymbol{\omega}$ postulated
in Eq. (\ref{eq:Postulate}). 

\section{$\mathcal{N}=4$ Supersymmetric Yang-Mills Theory\label{sec:-Supersymmetric-Yang-Mills}}

\subsection{Review}

The starting point of our analysis is the Parke-Taylor (PT) formula,
which provides an elegant representation for the tree-level scattering
amplitude of $n$ gluons in a MHV configuration. Originally proposed
by \citet{parke1986amplitude} and subsequently given a rigorous proof
by \citet{berends1988recursive}, this formula expresses the amplitude
in the helicity configuration $1^{-},2^{-},3^{+},...,n^{+}$ as\footnote{For modern pedagogical introductions, see \citet{elvang2013scattering,badger2024scattering}.}:
\begin{equation}
\mathcal{A}_{n}^{a_{1}...a_{n}}=ig^{n-2}\delta^{\left(4\right)}\left(\sum_{i=1}^{n}p_{i}^{\mu}\right)\left(PT\right)_{n}^{a_{1}...a_{n}},
\end{equation}
where:
\begin{equation}
\left(PT\right)_{n}^{a_{1}...a_{n}}=\frac{\left\langle \nu_{1},\nu_{2}\right\rangle ^{4}\text{Tr}\left(\mathsf{R}^{a_{1}}\mathsf{R}^{a_{n}}...\mathsf{R}^{a_{n}}\right)}{\left\langle \nu_{1},\nu_{2}\right\rangle \left\langle \nu_{2},\nu_{3}\right\rangle ...\left\langle \nu_{n},\nu_{1}\right\rangle }.
\end{equation}
Here, $g$ denotes the Yang-Mills coupling constant, while $\delta^{\left(4\right)}$
imposes the conservation of four-momentum across the scattering process.
We denote by $\mathsf{R}^{a}$ the generators of the adjoint representation
of the gauge group $SO(\tilde{N})$, with their normalisation specified
by $\text{Tr}\left(\mathsf{R}^{a}\mathsf{R}^{b}\right)=2\delta^{ab}$.
These generators satisfy the commutation relations $[\mathsf{R}^{a},\mathsf{R}^{b}]=iC^{abc}\mathsf{R}^{c}$,
where $\mathsf{C}^{abc}$ are the structure constants of $SO(\tilde{N})$.
\emph{For the sake of precision, we emphasise that the gauge group
under consideration in this section differs from that employed in
the preceding analysis.}

The scattering amplitude $\mathcal{A}_{n}^{a_{1}...a_{n}}$ can be
reformulated in terms of the gluon frequencies $s_{i}$ and the two-component
spinors $\pi_{i}^{A}\coloneqq\left(z_{i},1\right)^{T}$ as follows:
\begin{equation}
\mathcal{A}_{n}^{a_{1}...a_{n}}=ig^{n-2}\int\frac{d^{4}X}{\left(2\pi\right)^{4}}\prod_{i=1}^{n}s_{i}^{e_{i}}e^{is_{i}q\left(z_{i},\bar{z}_{i}\right)\cdot X}\frac{\left\langle \pi_{1},\pi_{2}\right\rangle ^{4}\text{Tr}\left(\mathsf{R}^{a_{1}}\mathsf{R}^{a_{n}}...\mathsf{R}^{a_{n}}\right)}{\left\langle \pi_{1},\pi_{2}\right\rangle \left\langle \pi_{2},\pi_{3}\right\rangle ...\left\langle \pi_{n},\pi_{1}\right\rangle }.
\end{equation}
In this representation, the exponents $e_{1}=e_{2}=1$ and $e_{3}=...=e_{n}=-1$
correspond to the MHV configuration $1^{-},2^{-},3^{+},...,n^{+}$.

\paragraph{Supersymmetry.}

The transition to $\mathcal{N}=4$ SYM theory is facilitated by the
introduction of Grassmann-valued two-component spinors $\theta^{A\alpha}$
$\left(\alpha=1,...,4\right)$, normalised as $\int d^{2}\theta\theta_{A}^{\alpha}\theta_{B}^{\alpha}=\varepsilon_{AB}$.
From these spinors, we define the coefficients:
\[
\zeta_{i}^{\alpha}\coloneqq\theta^{A\alpha}\pi_{iA}\,\,\,\left(i=1,...,n\right).
\]
This allows us to express the Berezin integral identity:
\begin{equation}
\int d^{8}\theta\prod_{\alpha=1}^{4}\zeta_{i}^{\alpha}\zeta_{j}^{\alpha}=\left\langle \pi_{i},\pi_{j}\right\rangle ^{4}.
\end{equation}
Thus, the amplitude $\mathcal{A}_{n}$ can be reformulated as:
\begin{equation}
\mathcal{A}_{n}^{a_{1}...a_{n}}=ig^{n-2}\int\frac{d^{4}X}{\left(2\pi\right)^{4}}\int d^{8}\theta\prod_{\alpha=1}^{4}\zeta_{1}^{\alpha}\zeta_{2}^{\alpha}\text{Tr}\left(\prod_{i=1}^{n}s_{i}^{e_{i}}e^{is_{i}q\left(z_{i},\bar{z}_{i}\right)\cdot X}\frac{\mathsf{R}^{a_{i}}}{\left\langle \pi_{i},\pi_{i+1}\right\rangle }\right).
\end{equation}

This reformulation provides a compelling motivation for introducing
the $\mathcal{N}=4$ superfield $Z_{i}$, defined as follows:
\begin{equation}
Z_{i}=a^{-}\left(z_{i},\bar{z}_{i}\right)+\zeta_{i}^{\alpha}\lambda_{\alpha}+\frac{1}{2}\zeta_{i}^{\alpha}\zeta_{i}^{\beta}\phi_{\alpha\beta}+\frac{1}{3!}\zeta_{i}^{\alpha}\zeta_{i}^{\beta}\zeta_{i}^{\gamma}\varepsilon_{\alpha\beta\gamma\delta}\tilde{\lambda}^{\delta}+a^{+}\left(z_{i},\bar{z}_{i}\right)\prod_{\alpha=1}^{4}\zeta_{i}^{\alpha},
\end{equation}
which describes the particle content of $\mathcal{N}=4$ SYM theory.
Here, the term $a_{i}^{-}\coloneqq a^{-}\left(z_{i},\bar{z}_{i}\right)$
corresponds to the vacuum expectation value of the annihilation operator
for a negative-helicity gluon, while $a_{i}^{+}\coloneqq a^{+}\left(z_{i},\bar{z}_{i}\right)$
represents the analogous quantity for a positive-helicity gluon. The
dependence of the superfield $Z_{i}$ on the celestial sphere coordinates
$\left(z_{i},\bar{z}_{i}\right)\in\mathbf{CP}^{1}$ arises through
the coefficients $\zeta_{i}^{\alpha}$.

Consequently, we introduce the (partial) super-amplitude as:
\begin{equation}
\mathbb{A}_{n}^{a_{1}...a_{n}}=ig^{n-2}\int\frac{d^{4}X}{\left(2\pi\right)^{4}}\int d^{8}\theta\text{Tr}\left(\prod_{i=1}^{n}s_{i}^{e_{i}}e^{is_{i}q\left(z_{i},\bar{z}_{i}\right)\cdot X}\frac{\mathsf{R}^{a_{i}}Z_{i}}{\left\langle \pi_{i},\pi_{i+1}\right\rangle }\right).
\end{equation}

\paragraph{Celestial Super-amplitude.}

The celestial (partial) super-amplitude $\widehat{\mathbb{A}}_{n}^{a_{1}...a_{n}}$
is subsequently defined as the $\varepsilon$-regulated Mellin transform:
\begin{equation}
\widehat{\mathbb{A}}_{n}^{a_{1}...a_{n}}\coloneqq\prod_{i=1}^{n}\int_{\left(0,\infty\right)}\frac{ds_{i}}{s_{i}}s_{i}^{\Delta_{i}}e^{-\varepsilon s_{i}}\mathbb{A}_{n}^{a_{1}...a_{n}}.
\end{equation}
By performing the integral transform, $\widehat{\mathbb{A}}_{n}^{a_{1}...a_{n}}$
can be expressed in terms of the celestial wavefunctions $\phi_{2h_{i}}\left(z_{i},\bar{z}_{i}\big|X\right)$
as:
\begin{equation}
\widehat{\mathbb{A}}_{n}^{a_{1}...a_{n}}=ig^{n-2}\int\frac{d^{4}X}{\left(2\pi\right)^{4}}\int d^{8}\theta\text{Tr}\left(\prod_{i=1}^{n}\phi_{2h_{i}}\left(z_{i},\bar{z}_{i}\big|X\right)\frac{\mathsf{R}^{a_{i}}Z_{i}}{\left\langle \pi_{i},\pi_{i+1}\right\rangle }\right),
\end{equation}
where the scaling dimensions for the chosen MHV configuration are
determined by the relations:
\begin{equation}
2h_{i}\coloneqq\Delta_{i}+e_{i},\,\,\,\forall\,1\leq i\leq n.
\end{equation}

By employing the formalism of celestial leaf amplitudes, and under
the simplifying assumption that all gluons are outgoing, the spacetime
integral appearing in the preceding expression can be evaluated as
follows:
\begin{equation}
\int d^{4}X\prod_{i=1}^{n}\phi_{2h_{i}}\left(z_{i},\bar{z}_{i}\big|X\right)=2\pi\delta\left(\beta\right)\int_{\mathbf{H}_{3}}d^{3}x\prod_{i=1}^{n}G_{2h_{i}}\left(z_{i},\bar{z}_{i}\big|x\right)+(\bar{z}_{i}\rightarrow-\bar{z}_{i}).
\end{equation}
Here, the quantity $\beta$ is defined as:
\begin{equation}
\beta\coloneqq2\sum_{i=1}^{n}h_{i}-4,
\end{equation}
which represents the total scaling dimension associated with the scattering
process in question. 

Consequently, the celestial super-amplitude can be represented in
the compact form:
\begin{equation}
\widehat{\mathbb{A}}_{n}^{a_{1}...a_{n}}=\frac{ig^{n-2}}{\left(2\pi\right)^{3}}\delta\left(\beta\right)A_{n}^{a_{1}...a_{n}}+\left(\bar{z}_{i}\rightarrow-\bar{z}_{i}\right),
\end{equation}
where the leaf amplitude for gluons $A_{n}^{a_{1}...a_{n}}$ is defined
by:
\begin{equation}
A_{n}^{a_{1}...a_{n}}\coloneqq\int_{\mathbf{H}_{3}}d^{3}x\int d^{8}\theta\,\text{Tr}\prod_{i=1}^{n}G_{2h_{i}}\left(z_{i},\bar{z}_{i}\big|x\right)\frac{\mathsf{R}^{a_{i}}Z_{i}}{\left\langle \pi_{i},\pi_{i+1}\right\rangle }.
\end{equation}

\begin{rem*}
Inspired by heuristic considerations analogous to those employed by
\citet{witten2004perturbative} in his formulation of twistor string
theory, we find that the gluon leaf $n$-amplitude $A_{n}^{a_{1}...a_{n}}$
derived above admits the following representation as an integral over
spinor variables:
\begin{equation}
A_{n}^{a_{1}...a_{n}}=\prod_{i=1}^{n}\int d^{2}\mu_{i}\frac{\Gamma\left(2h_{i}\right)}{\left(-i\mu_{i}\cdot\bar{\pi}_{i}\right)}\int_{\mathbf{H}_{3}}d^{3}x\int d^{8}\theta\,\phi^{a_{1}...a_{n}},
\end{equation}
where the mapping $(\mu_{\dot{A}i},\pi_{i}^{A})\in\text{\textbf{MT}}\mapsto\phi^{a_{1}...a_{n}}$
on the $\mathcal{N}=4$ supersymmetric extension of minitwistor space
$\mathbf{MT}$ (associated with Euclidean $AdS_{3}$) is defined as
follows: 
\begin{equation}
\phi^{a_{1}...a_{n}}=\text{Tr}\prod_{i=1}^{n}\delta^{\left(2\right)}\left(\mu_{\dot{A}i}-\pi_{i}^{A}x_{A\dot{A}}\right)\frac{\mathsf{R}^{a_{i}}Z_{i}(\theta^{A\alpha},\bar{\theta}_{\alpha}^{\dot{A}})}{\left\langle \pi_{i},\pi_{i+1}\right\rangle }.
\end{equation}
Here, the Grassmann-valued spinors $(\theta^{A\alpha},\bar{\theta}_{\alpha}^{\dot{A}})$
appear in the arguments of the superfield $Z_{i}$ solely to emphasise
how the dependence of $\phi^{a_{1}...a_{n}}$ on the supersymmetric
extension of $\mathbf{MT}$ is introduced. Adopting the terminology
of \citet{cachazo2013gravity,cachazo2014gravity,adamo2013twistor},
we interpret $\phi^{a_{1}...a_{n}}$ as the multi-gluon wavefunction
defined on minitwistor space. The set $\{(x^{\mu},\theta^{A\alpha},\bar{\theta}_{\alpha}^{\dot{A}},\mu_{\dot{A}},\pi^{A})\}$
then represents the $\mathcal{N}=4$ supersymmetric extension of the
configuration space. Note that this supersymmetric configuration space
is intimately related to the mathematical notion of \emph{graded manifolds},
which were originally introduced in the study of dynamical systems
possessing fermionic degrees of freedom. We refer the reader to the
works of \citet{kostant2006graded}, \citet{berezin2013introduction}
and \citet{leites1980introduction}. 
\end{rem*}

\subsection{Euclidean $AdS_{3}$ String Theory and Gluon Leaf Amplitudes}

Applying an analogous line of reasoning to that employed in the preceding
section, we establish a correspondence between the gluon leaf amplitude
$A_{n}$ and (Euclidean) $AdS_{3}$ string theory based on the following
considerations. Recall from the prior discussion that, in the minisuperspace
limit $k\rightarrow\infty$, the correlation functions of the primary
fields $\Phi^{h_{i}}\left(w_{i}\big|z_{i}\right)$ in the level-$k$
$H_{3}^{+}$-WZNW model reduce to a contact Feynman-Witten diagram
describing the propagation of massless particles on the hyperboloid
$\mathbf{H}_{3}$. Explicitly, this is expressed as:
\begin{equation}
\lim_{k\rightarrow\infty}\prod_{i=1}^{n}\int d^{2}w_{i}\,\Gamma\left(2h_{i}\right)\langle\Phi^{h_{1}}\left(w_{1}\big|z_{1}\right)...\Phi^{h_{n}}\left(w_{n}\big|z_{n}\right)\rangle=\int_{\mathbf{H}_{3}}d^{3}x\,\prod_{i=1}^{n}G_{2h_{i}}\left(z_{i},\bar{z}_{i}\big|x\right).
\end{equation}
Here, $G_{2h_{i}}\left(z_{i},\bar{z}_{i}\big|x\right)$ denotes the
bulk-to-boundary Green's function associated with the covariant Laplacian
on $\mathbf{H}_{3}$.

We now consider a solution to the equations of motion of (Euclidean)
$AdS_{3}$ string theory, defined on the target space $AdS_{3}\times\tilde{X}$,
where $\tilde{X}$ is a compact manifold. Let $K^{a}\left(z\right)$
denote a level-one $SO(\tilde{N})$ WZNW current algebra defined on
the boundary of (Euclidean) $AdS_{3}$, which is identified with the
celestial sphere. We postulate that the currents $K^{a}\left(z\right)$
arises from the CFT defined on $\tilde{X}$. 

Drawing upon the framework established by \citet{giveon1998comments},
\citet{dolan1999vertex} and \citet{dolan2001vertex}, it is known
that, given a spinless operator $W_{\tilde{X}}$ in the $\tilde{X}$-CFT,
physical vertex operators can be constructed in the form $W_{\tilde{X}}V_{jm\bar{m}}$,
where $V_{jm\bar{m}}$ corresponds to the worldsheet vertex operators. 

Building on this formalism, we define the \emph{celestial gluon vertex
operators} as:
\begin{equation}
\mathcal{U}_{i}^{a_{i}}\coloneqq\int d^{2}w_{i}\Gamma\left(2h_{i}\right)K^{a_{i}}(z_{i})\Phi^{h_{i}}\left(w_{i}\big|z_{i}\right).
\end{equation}
The gluon leaf amplitude can thus be derived from the mini-superspace
limit $\left(k\rightarrow\infty\right)$ of Euclidean $AdS_{3}$ string
theory and is expressed as:
\begin{equation}
A_{n}^{a_{1}...a_{n}}=\lim_{k\rightarrow\infty}\int d^{8}\theta\langle\mathcal{U}_{1}^{a_{1}}...\mathcal{U}_{n}^{a_{n}}\rangle.
\end{equation}

\begin{rem*}
It is worth emphasising that the correspondence between celestial
CFT and Euclidean $AdS_{3}$ string theory presented herein constitutes
a significant refinement over the holographic dictionary proposed
in our earlier work (\citet{mol2024comments}). In the current formulation,
celestial vertex operators for gluons and gravitons are treated on
an equal footing. Specifically, both are constructed exclusively from
quantities arising as solutions to the equations of motion of $AdS_{3}$
string theory. The distinction between these operators lies solely
in their respective numbers of supersymmetry generators and the associated
gauge groups.
\end{rem*}

\subsection{Generating Functional}

Employing reasoning analogous to that utilised in the preceding section,
we construct a generating functional for the gluon leaf amplitudes,
$A_{n}^{a_{1}...a_{n}}$, based on the chiral determinant representation
of the WZNW action. The main distinction between the $\mathcal{N}=8$
Supergravity and $\mathcal{N}=4$ SYM cases lies in the number of
supersymmetry generators and the structure of the respective gauge
groups.

For the sake of clarity, we present a self-contained derivation rather
than merely extrapolating the gravitational results to the gauge-theoretic
case. While this approach entails some repetition, it serves to highlight
the features of the gauge theory case. Nonetheless, it will become
apparent that the fundamental structure of the generating functional
remains invariant across both cases. This ``universality'' enables
a systematic derivation of the action integral describing the proposed
holographic $AdS_{3}$ theory.

Let $\tilde{\mathcal{S}}[G]$ denote the action functional for the
level-one $SO(\tilde{N})$ WZNW model defined on $\mathbf{CP}^{1}$,
where the dynamical variable is the matrix field $G=G(z,\bar{z})$,
representing a mapping from $\mathbf{CP}^{1}$ to the Lie group $SO(\tilde{N})$.
From $G$, we construct the differential one-form: 
\begin{equation}
\mathscr{A}\coloneqq\mathscr{A}_{z}dz+\mathscr{A}_{\bar{z}}d\bar{z},\label{eq:Differential-1}
\end{equation}
on $\mathbf{CP}^{1}$, with its holomorphic and anti-holomorphic components
defined as follows:
\begin{equation}
\mathscr{A}_{z}=-(\partial_{z}G)G^{-1},\,\,\,\mathscr{A}_{\bar{z}}=-(\partial_{\bar{z}}G)G^{-1}.
\end{equation}
If $G$ satisfies the equations of motion of the WZNW model, the associated
one-form $\mathscr{A}$ satisfies the ``zero-curvature'' condition:
\begin{equation}
\mathcal{F}[\mathscr{A}]\coloneqq\partial_{z}\mathscr{A}_{\bar{z}}-\partial_{\bar{z}}\mathscr{A}_{z}+[\mathscr{A}_{z},\mathscr{A}_{\bar{z}}]=0.
\end{equation}
Observe the analogy between this equality and the ``flatness'' condition
$\mathcal{F}[\boldsymbol{\omega}]=0$ encountered in the gravitational
context. This parallel naturally suggests an interpretation of the
differential one-form $\mathscr{A}$ as an $\mathfrak{so}(\tilde{N})$-valued
connection one-form on $\mathbf{CP}^{1}$, associated with a gauge
potential whose field strength is characterised by the curvature two-form
$\mathcal{F}[\mathscr{A}]$. Consequently, we define the holomorphic
component of the gauge-covariant derivative operator associated to
the differential one-form $\mathscr{A}$ as:
\begin{equation}
\text{\b{d}}_{z}\coloneqq\partial_{z}+[\mathscr{A}_{z},\cdot].
\end{equation}

As reviewed by \citet{nair2005chern}, the action functional $\tilde{\mathcal{S}[G]}$
for the level-one $SO(\tilde{N})$ WZNW model on $\mathbf{CP}^{1}$
can be expressed in terms of the chiral Dirac determinant as follows:
\begin{equation}
\mathcal{\tilde{\mathcal{S}}}[G]\eqqcolon\mathcal{\tilde{\mathcal{S}}}[\mathscr{A}]=\text{Tr}\log\text{\b{d}}_{z}-\text{Tr}\log\partial_{z}.
\end{equation}
This expression admits a formal expansion in terms of the components
$\mathscr{A}_{z}^{a_{i}}\left(z_{i},\bar{z}_{i}\right)$:
\[
\tilde{\mathcal{S}}[\mathscr{A}]=\sum_{m\geq2}\frac{\left(-1\right)^{m+1}}{m}\int\frac{d^{2}z_{1}}{\pi}...\int\frac{d^{2}z_{m}}{\pi}\text{Tr}\left(\frac{\mathscr{A}_{z}^{a_{1}}\left(z_{1},\bar{z}_{1}\right)...\mathscr{A}_{z}^{a_{n}}\left(z_{m},\bar{z}_{m}\right)}{z_{12}z_{23}...z_{m1}}\right),
\]
where $a_{i}$ represents the colour indices associated with the $SO(\tilde{N})$
gauge group. The primary distinction between this formulation and
the analogous expression for $\mathcal{S}[\boldsymbol{\omega}]$ in
the gravitational case lies in the structure of the colour indices.

As in the preceding analysis, we consider the possibility of lifting
the gauge potential $\mathscr{A}$ from $\mathbf{CP}^{1}$ to $\tilde{\mathscr{A}}$
defined on the $\mathcal{N}=4$ supersymmetric extension of $\mathbf{H}_{3}$.
The holomorphic component of the extended potential, $\tilde{\mathscr{A}}_{z}^{a_{i}}$,
is expressed as:
\begin{equation}
\tilde{\mathscr{A}}_{z}^{a_{i}}\left(z_{i},\bar{z}_{i}\right)=\int_{\mathcal{P}}\frac{d\Delta_{i}}{2\pi i}G_{2h_{i}}\left(z_{i},\bar{z}_{i}\big|x\right)\mathsf{R}^{a_{i}}Z_{i}.\label{eq:Postulate-1}
\end{equation}
Here, $\tilde{\mathscr{A}}_{z}^{a_{i}}$ differs from $\boldsymbol{\Omega}_{i}$,
introduced in the gravitational case, in two key respects: the number
of super-symmetries, contained in the superfield $Z_{i}$, and the
gauge structure, described by the $\mathsf{R}^{a}$ generators. Consequently,
we define the ``effective'' action functional for the gluon leaf amplitudes
as follows:
\begin{equation}
\exp(\mathcal{W}_{YM}[\tilde{\mathscr{A}}])\coloneqq\int_{\mathbf{H}_{3}}d^{3}x\int d^{8}\theta\tilde{\mathcal{S}}[\tilde{\mathscr{A}}_{z}^{a_{i}}].
\end{equation}

By taking functional derivatives of $e^{\mathcal{W}_{YM}}$ with respect
to $a_{i}^{\pm a_{i}}$, representing the vacuum expectation values
for the annihilation operators of gluons with positive and negative
helicities, we obtain the gluon leaf amplitude:
\begin{equation}
A_{n}=\frac{\delta}{\delta a_{1}^{-}}\frac{\delta}{\delta a_{2}^{-}}\frac{\delta}{\delta a_{3}^{+}}...\frac{\delta}{\delta a_{n}^{+}}\exp(\mathcal{W}_{YM}[\tilde{\mathscr{A}}])\Big|_{\tilde{\mathscr{A}}=0}.\label{eq:Generating-Functional-1}
\end{equation}

This construction demonstrates that the ``effective'' action $e^{\mathcal{W}_{YM}}$
serves as the generating functional for the gluon leaf amplitudes,
as previously asserted. Furthermore, as noted at the outset of this
section, the principal differences between the gravitational and gauge-theoretic
cases arises from the number of degrees of freedom contained in the
respective superfields ($\Psi_{i}$ \emph{versus} $Z_{i}$) and the
structure of the Lie group ($\mathsf{T}^{a}$ generators \emph{versus
$\mathsf{R}^{a}$} generators).

Despite these distinctions, the fundamental structure of the generating
functional remains invariant. This ``universality'' enables a systematic
and unified analysis of the action functional governing the holographic
$AdS_{3}$ theory for both gravitons and gluons. 

\section{Holographic $AdS_{3}$ Models}

\subsection{Introductory Remarks}

To explain our approach to deriving the action integral for the WZNW-like
field theory on $AdS_{3}$ that generates the leaf amplitudes for
gravitons and gluons, we begin by briefly reviewing, in accordance
with \citet{weinberg2005quantum}, the formulation of the $S$-matrix
in terms of the effective action.

Let $\varphi^{I}(x)$ denote a set of fields indexed by $I$, and
let $\Gamma[\varphi^{I}(x)]$ represent the effective quantum action.
The quantum equations of motion are determined by the stationarity
points of $\Gamma[\varphi^{I}(x)]$, satisfying the condition:
\begin{equation}
\frac{\delta}{\delta\varphi^{I}}\Gamma=0.
\end{equation}
The generating functional that yields the $S$-matrix elements is
then defined as:
\begin{equation}
F\coloneqq\exp(i\Gamma[\varphi^{I}])\Big|_{\frac{\delta\Gamma}{\delta\varphi^{I}}=0}.
\end{equation}
In perturbative quantum field theory, the solutions to the equations
of motion are expanded around the free field configuration:
\begin{equation}
\varphi^{I}(x)\propto\int dk\,[a(k)u_{k}(x)+a^{*}(k)u_{k}^{*}(k)],
\end{equation}
where $u_{k}(x)$ are plane-wave modes, and $a(k)$, $a^{*}(k)$ denote
the mode coefficients. The scattering amplitude for a process of the
form $p_{1},p_{2}...p_{n}\rightarrow p'_{1},p'_{2}...p'_{m}$ is computed
using the expression:
\begin{equation}
\mathcal{A}=\left(\prod_{i=1}^{n}\frac{\delta}{\delta a(p_{i})}\prod_{j=1}^{n}\frac{\delta}{\delta a(p'_{j})}F\right)_{a(k)=a^{*}(k)=0}.\label{eq:Example}
\end{equation}

This expression coincides formally with the graviton leaf amplitude
$M_{n}$ and the gluon leaf amplitude $A_{n}^{a_{1}...a_{n}}$ obtained
in Eqs. (\ref{eq:Generating-Functional}) and (\ref{eq:Generating-Functional-1}).
Therefore, the next level of abstraction is introduced as follows:
we \emph{postulate} that the classical \emph{effective} actions $\mathcal{W}_{SUGRA}[\boldsymbol{\Omega}]$
and $\mathcal{W}_{YM}[\tilde{\mathscr{A}}]$, which govern the holographic
$AdS_{3}$ theories, take the following forms:
\begin{equation}
\mathcal{W}_{SUGRA}[\boldsymbol{\Omega}]=\int_{\mathbf{H}_{3}}d^{3}x\int d^{16}\theta\,\mathcal{S}[\boldsymbol{\Omega}],
\end{equation}
and:
\begin{equation}
\mathcal{W}_{YM}[\tilde{\mathscr{A}}]\coloneqq\int_{\mathbf{H}_{3}}d^{3}x\int d^{8}\theta\tilde{\mathcal{S}}[\tilde{\mathscr{A}}_{z}^{a_{i}}],
\end{equation}
which correspond, respectively, to the dual descriptions of the MHV
subsector of celestial CFT for $\mathcal{N}=8$ Supergravity and $\mathcal{N}=4$
SYM theory. 

Under these assumptions, the results derived in Sections II and III,
specifically Eqs. (\ref{eq:Generating-Functional}) and (\ref{eq:Generating-Functional-1}),
are precisely reproduced as particular instances of Eq. (\ref{eq:Example})
for the respective actions. Explicitly, these amplitudes are expressed
as:
\begin{equation}
M_{n}=\left(\prod_{i=1}^{n}\frac{\delta}{\delta h^{\sigma_{i}}\left(z_{i},\bar{z}_{i}\right)}F_{SUGRA}\right)_{\boldsymbol{\Omega}=0},\label{eq:Generating-Functional-2}
\end{equation}
and:
\begin{equation}
A_{n}=\left(\prod_{i=1}^{n}\frac{\delta}{\delta a^{\bar{\sigma}_{i}}\left(z_{i},\bar{z}_{i}\right)}F_{YM}\right)_{\tilde{\mathscr{A}}=0},\label{eq:Generating-Functional-1-1}
\end{equation}
where the generating functionals are defined as:
\begin{equation}
F_{SUGRA}\coloneqq\exp(\mathcal{W}_{SUGRA}[\boldsymbol{\Omega}]),
\end{equation}
and:
\begin{equation}
F_{YM}\coloneqq\exp(\mathcal{W}_{YM}[\tilde{\mathscr{A}}]).
\end{equation}
In these formulae, $\sigma_{i}$ and $\bar{\sigma}_{i}$ denote the
helicities of gravitons and gluons, respectively, while $h^{\sigma_{i}}\left(z_{i},\bar{z}_{i}\right)$
and $a^{\bar{\sigma}_{i}}\left(z_{i},\bar{z}_{i}\right)$ represent
the expectation values of the annihilation operators for gravitons
and gluons. These operators are contained within the superfields $\Psi_{i}$
(for $\mathcal{N}=8$ Supergravity) and $Z_{i}$ (for $\mathcal{N}=4$
SYM theory), respectively.

We now delineate the framework for constructing holographic $AdS_{3}$
theories. \citet{abe2005multigluon} and \citet{abe5analysis} demonstrated
that the superspace constraints of $\mathcal{N}=4$ SYM theory and
the anti-self-dual sector of $\mathcal{N}=8$ Supergravity can be
embedded in a supersymmetric extension of twistor space. Within this
formalism, it was further shown by these authors that one can impose
the so-called \emph{chiral semi-analytic gauge} on the constraint
equations, adopting the terminology introduced in the harmonic superspace
literature, as comprehensively revised by \citet{galperin2001harmonic}.
From these results, we extract two important corollaries that inform
our subsequent analysis.

The first corollary asserts that, upon imposing the chiral semi-analytic
gauge, the residual constraint is precisely identified with the zero-curvature
conditions $\mathcal{F}[\boldsymbol{\omega}]=0$ and $\mathcal{F}[\mathscr{A}]=0$,
as derived in Sections II and III for $\mathcal{N}=8$ Supergravity
and $\mathcal{N}=4$ SYM theory, respectively. These zero-curvature
conditions will emerge as the Euler-Lagrange equations of the proposed
action for the holographic $AdS_{3}$ models.

The second corollary pertains to the geometric interpretation of the
celestial leaf amplitudes. As established in Sections II and III,
multi-graviton and multi-gluon celestial leaf amplitudes naturally
admit a representation within minitwistor space. Consequently, solving
the so-called \emph{analyticity} \emph{condition} for the gauge superpotentials
in the supersymmetric twistor space, and subsequently performing a
scaling reduction from twistor to minitwistor space, following the
approach of \citet{bu2023celestial}, yields the lifted potentials
$\boldsymbol{\Omega}$ and $\tilde{\mathscr{A}}$. These potentials
were introduced in Eqs. (\ref{eq:Postulate}) and (\ref{eq:Postulate-1})
of Sections II and III, respectively, and are essential for constructing
the generating functionals for celestial leaf amplitudes. For completeness,
we recall that the holomorphic components of these potentials are
explicitly given by:
\begin{equation}
\Omega_{z}(z_{i},\bar{z}_{i}\big|x,\theta,\bar{\theta})\coloneqq\int_{\mathcal{P}}\frac{d\Delta_{i}}{2\pi i}G_{2h_{i}}\left(z_{i},\bar{z}_{i}\big|x\right)\Psi_{i},
\end{equation}
and:
\begin{equation}
\tilde{\mathscr{A}}_{z}^{a_{i}}\left(z_{i},\bar{z}_{i}\right)=\int_{\mathcal{P}}\frac{d\Delta_{i}}{2\pi i}G_{2h_{i}}\left(z_{i},\bar{z}_{i}\big|x\right)\mathsf{R}^{a_{i}}Z_{i}.
\end{equation}

Finally, we will propose an action integral for the holographic WZNW-like
field theories on $AdS_{3}$, demonstrate that its Euler-Lagrange
equations recover the superspace constraints in the chiral semi-analytic
gauge (or equivalently, the zero-curvature conditions, $\mathcal{F}[\boldsymbol{\omega}]=0$
and $\mathcal{F}[\mathscr{A}]=0$) and conclude by showing that the
corresponding on-shell effective actions yield Eqs. (\ref{eq:Generating-Functional})
and (\ref{eq:Generating-Functional-1}).

\subsection{Holomorphic Superpotential\label{subsec:Effective-Potential}}

For the purposes of our present analysis, the main lessons from \citet{abe2005multigluon}
and \citet{abe5analysis} can be summarised as follows. The superspaces
associated with $\mathcal{N}=4$ SYM theory and the anti-self-dual
sector of $\mathcal{N}=8$ Supergravity may be extended to include
a pair of two-component spinors, $\left(\mu^{A},\nu_{\dot{A}}\right)$,
together with an auxiliary frame vector $W_{A\dot{A}}$, related through
the equation:
\begin{equation}
\nu_{\dot{A}}=\mu^{A}W_{A\dot{A}}.
\end{equation}

In what follows, this extension of superspace will be interpreted
\emph{heuristically} as a supersymmetric generalisation of twistor
space. For a mathematically rigorous account of the construction of
twistor superspace, the reader is referred to the original works of
\citet{Lukierski:1986kg,Lukierski:1988vw,Kotrla:1984ky}. For recent
developments, see \citet{Popov:2004rb} and \citet{Wolf:2006me}.

Within this framework, \citet{abe2005multigluon} and \citet{abe5analysis}
established that the constraints of $\mathcal{N}=4$ SYM theory and
the anti-self-dual sector of $\mathcal{N}=8$ Supergravity can be
formulated in the so-called \emph{chiral semi-analytic gauge}, a terminology
originating in the harmonic superspace literature, as reviewed by
\citet{galperin2001harmonic,ivanov2023n}. In this gauge, the superpotentials,
denoted $H^{++}$ and $H^{--}$, satisfy the following system of constraint
equations:
\begin{equation}
\overline{D}_{\dot{A}}^{\alpha}H^{++}=0,\,\,\,\text{(chirality condition)}\label{eq:Chirality}
\end{equation}
\begin{equation}
D_{\alpha}^{(+)}H^{++}=0\,\,\,\text{(analyticity condition)},\label{eq:Constraint}
\end{equation}
\begin{equation}
D^{(++)}H^{--}-D^{(--)}H^{++}+[H^{++},H^{--}]=0,\label{eq:Constraint-1}
\end{equation}
where the derivative operators are defined as follows:
\begin{equation}
D_{A\alpha}=\frac{\partial}{\partial\theta^{A\alpha}}+i(\sigma^{\mu})_{A\dot{A}}\bar{\theta}_{\alpha}^{\dot{A}}\frac{\partial}{\partial x^{\mu}}\,\,\,\overline{D}_{\dot{A}}^{\alpha}\coloneqq-\frac{\partial}{\partial\bar{\theta}_{\alpha}^{\dot{A}}}-i\theta^{A\alpha}(\sigma^{\mu})_{A\dot{A}}\frac{\partial}{\partial x^{\mu}},\label{eq:Definition-1}
\end{equation}
\begin{equation}
D_{\alpha}^{(+)}\coloneqq\mu^{A}D_{A\alpha},\,\,\,D^{(++)}\coloneqq\mu^{A}\frac{\partial}{\partial\bar{\nu}^{A}},\,\,\,D^{(--)}\coloneqq-\bar{\nu}^{A}\frac{\partial}{\partial\mu^{A}}.\label{eq:Definition-2}
\end{equation}
A concise definition of the gauge superpotentials $H^{++}$ and $H^{--}$
is provided by \citet{abe2005multigluon} and \citet{abe5analysis}.
For a rigorous mathematical construction of $H^{++}$ and $H^{--}$,
and the ``harmonic derivatives'' $D_{\alpha}^{(+)}$, $D^{(++)}$
and $D^{(--)}$, as well as their gauge-theoretic properties, we refer
the reader to \citet{galperin2001harmonic}.

To summarise, the superspace constraints of $\mathcal{N}=4$ SYM theory
and the anti-self-dual sector of $\mathcal{N}=8$ Supergravity (which,
as defined by \citet{abe5analysis}, is the sector of vanishing spinorial
curvature, $R_{\alpha\beta}^{ab}=0$), when expressed in the chiral
semi-analytic gauge within this twistor-like extension of superspace,
exhibit an \emph{identical structural form}. The above constraint
equations for these theories differ solely in the number of \emph{supersymmetry
generators} associated with the superpotentials and their respective\emph{
gauge groups}. Accordingly, we employ the notation $H^{++},H^{--}$
to refer generically to the gauge superpotentials for either of these
theories. In what follows, we adopt an abstract and generic approach,
ensuring that our results are applicable to both $\mathcal{N}=4$
SYM theory and the anti-self-dual sector of $\mathcal{N}=8$ Supergravity. 

\subsubsection{Local Parametrisation}

We now proceed to reformulate Eq. (\ref{eq:Constraint-1}) by introducing
a local parametrisation for the pair $(\mu^{A},\nu_{\dot{A}})$ of
two-component spinors, alongside the four-vector $W_{A\dot{A}}$.
Let us write $\mu^{A}=(\xi,\zeta)^{T}$ and assume the existence of
an open neighbourhood $\mathcal{U}$ where $\zeta\neq0$. In this
region, the local coordinate representation of $\mu^{A}$ on the complex
projective line can be defined by $z\coloneqq\xi/\zeta$. Furthermore,
we choose a convenient frame wherein the four-vector is fixed to $W_{A\dot{A}}=\delta_{A\dot{A}}$.
Under this parametrisation, the spinors $\mu^{A}$ and $\nu_{\dot{A}}$
can be identified, respectively, with $\mu^{A}=\pi^{A}=(z,1)^{T}$
and $\nu_{\dot{A}}=\bar{\pi}_{\dot{A}}=(\bar{z},-1)^{T}$ (up to little-group
rescaling), which were previously employed in the analysis of graviton
and gluon celestial amplitudes in Sections II and III. 

We now introduce a local parametrisation of the gauge superpotentials,
defined on the neighbourhood $\mathcal{U}$ as follows:
\begin{equation}
(H^{(++)})_{\mathcal{U}}=\xi\bar{\xi}^{-1}(1+z\bar{z})H_{z},\,\,\,(H^{(--)})_{\mathcal{U}}=\bar{\xi}\xi^{-1}(1+z\bar{z})H_{\bar{z}}.
\end{equation}
In this representation, the constraint expressed by Eq. (\ref{eq:Constraint-1})
assumes the form:
\begin{equation}
\partial_{z}H_{\bar{z}}-\partial_{\bar{z}}H_{z}+[H_{z},H_{\bar{z}}]=0.\label{eq:Constraint-2}
\end{equation}
This equation is thereby identified with the zero-curvature conditions
$\mathcal{F}[\boldsymbol{\omega}]=0$ and $\mathcal{F}[\mathscr{A}]=0$,
previously derived in Sections II and III in the contexts of $\mathcal{N}=8$
Supergravity and $\mathcal{N}=4$ SYM theory, respectively.

Henceforth, we \emph{formally} define $H\coloneqq H_{z}dz+H_{\bar{z}}d\bar{z}$
and designate $H$ as \emph{the} superpotential, with its holomorphic
and anti-holomorphic components represented by $H_{z}$ and $H_{\bar{z}}$,
respectively. Observe the structural analogy between the superpotential
$H$ thus defined and the $\mathbf{CP}^{1}$ differential one-forms
$\boldsymbol{\omega}=\omega_{z}dz+\omega_{\bar{z}}d\bar{z}$ (cf.
Eq. (\ref{eq:Differential})) and $\mathscr{A}=\mathscr{A}_{z}dz+\mathscr{A}_{\bar{z}}d\bar{z}$
(cf. Eq. (\ref{eq:Differential-1})), as introduced in Sections II
and III, respectively. Our objective is to show that this correspondence
\emph{extends beyond a merely formal analogy.} 

\subsubsection{Fourier Representation and Analyticity}

To address the analyticity condition, Eq. (\ref{eq:Constraint}),
we first examine the implications of the chirality constraint, Eq.
(\ref{eq:Chirality}), on the Fourier representation of the holomorphic
component $H_{z}$. This constraint necessitates that $H$ depends
on the Grassmann-valued two-component spinors $\theta^{A\alpha}$
and $\bar{\theta}_{\alpha}^{\dot{A}}$ exclusively through the combination:
\begin{equation}
Y^{\mu}(\mathbb{X})\coloneqq X^{\mu}+i\theta^{A\alpha}(\sigma^{\mu})_{A\dot{A}}\bar{\theta}_{\alpha}^{\dot{A}},
\end{equation}
where $\mathbb{X}^{I}\coloneqq(X^{\mu},\theta^{A\alpha},\bar{\theta}_{\alpha}^{\dot{A}})$
denotes the coordinates of the $\mathcal{N}$-extended supersymmetric
flat space.

Following \citet{weinberg2005quantum}, the holomorphic component
$H_{z}=H_{z}(\mathbb{X})$ of the superpotential is then expressed
in terms of an on-shell Fourier decomposition involving the mode functions
$\boldsymbol{\alpha}_{\sigma}^{a}(p)$:
\begin{equation}
H_{z}^{a}=\sum_{\sigma}\int d^{4}p\,\,\,\theta\left(p^{0}\right)\delta(p^{2})e^{ip\cdot Y}\boldsymbol{\alpha}_{\sigma}^{a}(p).\label{eq:Fourier}
\end{equation}
Here, $\sigma$ represents the helicity of the one-particle irreducible
representations of the Lorentz group, associated with the massless
gauge bosons of the theory (cf. \citet{weinberg1995quantum}), while
$a$ denotes the colour index corresponding to the gauge group. For
notational economy, we restrict our attention to the positive-frequency
modes, enforced through $\theta(p^{0})$, noting that the extension
to include negative-frequency modes is straighforward.

\paragraph{Momentum-space Parametrisation.}

As the Fourier decomposition in Eq. (\ref{eq:Fourier}) is explicitly
on-shell, enforced by the delta function $\delta(p^{2})$, the integration
is restricted to the null cone in momentum space, $\{p^{\mu}\big|p^{2}=0\}$.
To facilitate further analysis, we perform a change of integration
variables, mapping from Cartesian momentum-space coordinates $p^{\mu}$
to a parametrisation in terms of the frequency $s$ and celestial
sphere coordinates $z,\bar{z}$, given by:
\begin{equation}
p^{\mu}(s,z,\bar{z})=sq^{\mu}\left(z,\bar{z}\right)=s\left(1+z\bar{z},z+\bar{z},i\left(\bar{z}-z\right),1-z\bar{z}\right).
\end{equation}
In this parametrisation, the Lorentz-invariant measure transforms
as:
\begin{equation}
\frac{d^{3}\vec{p}}{2p^{0}}=dsd^{2}z\,is.
\end{equation}
Substituting this change of variables, the expression for $H_{z}^{a}$
becomes:
\begin{equation}
H_{z}^{a}=\sum_{\sigma}\int_{\mathbf{R}_{+}^{\times}}\frac{ds}{s}\int_{\mathbf{C}}d^{2}z\,s^{2}e^{isq(z,\bar{z})\cdot Y(\mathbb{X})}\boldsymbol{\alpha}_{\sigma}^{a}(sq(z,\bar{z})).
\end{equation}
Here, $\mathbf{R}_{+}^{\times}$ denotes the multiplicative group
of positive reals, with $ds/s$ representing the corresponding Haar
measure.

\paragraph{The Holomorphic Potential Density.}

Following \citet{pasterski2017conformal}, we recall the Mellin-integral
identity:
\begin{equation}
e^{isq\left(z,\bar{z}\right)\cdot Y\left(\mathbb{X}\right)}=\int_{\mathcal{P}}\frac{d\Delta}{2\pi i}s^{-\Delta}\frac{\Gamma\left(\Delta\right)}{\left(\varepsilon-iq\left(z,\bar{z}\right)\cdot Y\right)^{\Delta}},
\end{equation}
where $\mathcal{P}\coloneqq1+i\mathbf{R}$. Define the Mellin-transformed
Fourier modes $\hat{\boldsymbol{\alpha}}_{\sigma}^{a}=\hat{\boldsymbol{\alpha}}_{\sigma}^{a}(\Delta,z,\bar{z})$
by:
\begin{equation}
\hat{\boldsymbol{\alpha}}_{\sigma}^{a}\coloneqq i\int_{\mathbf{R}_{+}^{\times}}\frac{ds}{s}\,\,\,s^{2-\Delta}\boldsymbol{\alpha}_{\sigma}^{a}(sq(z,\bar{z})).
\end{equation}
With this definition, the holomorphic component of the superpotential
$H_{z}^{a}$ can equivalently be expressed as:
\begin{equation}
H_{z}^{a}=\sum_{\sigma}\int_{\mathbf{C}}d^{2}z\,h_{\sigma}^{a}(z,\bar{z}\big|\mathbb{X}),
\end{equation}
where the \emph{holomorphic superpotential density} $h_{\sigma}^{a}=h_{\sigma}^{a}(z,\bar{z}\big|\mathbb{X})$
is given by:
\begin{equation}
h_{\sigma}^{a}\left(z,\bar{z}\big|\mathbb{X}\right)\coloneqq\int_{\mathcal{P}}\frac{d\Delta}{2\pi i}\tilde{\phi}_{\Delta}(z,\bar{z}\big|\mathbb{X})\hat{\boldsymbol{\alpha}}_{\sigma}^{a}(\Delta,z,\bar{z}).
\end{equation}

Here, $\tilde{\phi}_{\Delta}$ represents the chiral supersymmetric
extension of the celestial conformal primary basis for massless scalars,
defined as:
\begin{equation}
\tilde{\phi}_{\Delta}(z,\bar{z}\big|\mathbb{X})\coloneqq\frac{\Gamma(\Delta)}{\left(\varepsilon-iq\left(z,\bar{z}\right)\cdot X+\xi^{\alpha}\bar{\xi}_{\alpha}\right)},
\end{equation}
where the coefficients:
\begin{equation}
\xi^{\alpha}\coloneqq\theta_{A}^{\alpha}\pi^{A},\,\,\,\bar{\xi}_{\alpha}\coloneqq\bar{\theta}_{\alpha}^{\dot{A}}\bar{\pi}_{\dot{A}}\,\,\,\left(\alpha=1,...,\mathcal{N}\right),
\end{equation}
are as defined in Sections II and III. Note that $h_{\sigma}^{a}$
is a density on the celestial sphere.

Employing the definitions of the superspace spinor and harmonic derivatives
from Eqs. (\ref{eq:Definition-1}) and (\ref{eq:Definition-2}), it
follows that $\tilde{\phi}_{\Delta}$ satisfies the following important
properties:
\begin{equation}
\overline{D}_{\dot{A}}^{\alpha}\tilde{\phi}_{\Delta}(z,\bar{z}\big|\mathbb{X})=0\,\,\,\text{(chirality condition),}
\end{equation}
\begin{equation}
D_{\alpha}^{(+)}\tilde{\phi}_{\Delta}(z,\bar{z}\big|\mathbb{X})=0\,\,\,\text{(analyticity condition).}\label{eq:Step}
\end{equation}

\paragraph{Analyticity.}

We now derive the implications of the analyticity constraint imposed
on the superpotential. Utilising Eq. (\ref{eq:Step}), together with
the Fourier representation, we find:
\begin{equation}
D_{i}^{(+)}h_{\sigma}^{a}=\int_{\mathscr{C}}\frac{d\Delta}{2\pi i}\tilde{\phi}_{\Delta}(z,\bar{z}\big|\mathbb{X})\pi^{A}\frac{\partial\hat{\boldsymbol{\alpha}}_{\sigma}^{a}(\Delta,z,\bar{z})}{\partial\theta^{Ai}}.
\end{equation}
The analyticity condition on the superpotential then requires that:
\begin{equation}
\pi^{A}\frac{\partial\hat{\boldsymbol{\alpha}}_{\sigma}^{a}(\Delta,z,\bar{z})}{\partial\theta^{Ai}}=0.
\end{equation}
Since $\hat{\boldsymbol{\alpha}}_{\sigma}^{a}$ depends on $\theta^{A\alpha}$
exclusively through the coefficients $\xi^{\alpha}$, it follows that
the Mellin-transformed mode functions must be of the form:
\begin{equation}
\hat{\boldsymbol{\alpha}}_{\sigma}^{a}(\Delta,z,\bar{z})=W_{\sigma,\Delta}(z,\bar{z})\mathsf{S}^{a}=\left(\alpha_{-}+\xi^{\alpha}\lambda_{\alpha}+...+\alpha_{+}\prod_{\alpha=1}^{\mathcal{N}}\xi^{\alpha}\right)\mathsf{S}^{a},
\end{equation}
where $W_{\sigma}$ is the superfield describing the particle multiplet
of the theory under consideration. Here, $\alpha_{-}=\alpha_{-}(\Delta,z,\bar{z})$
denotes the Mellin-transformed expectation value of the annihilation
operator for a massless boson with negative helicity, whereas $\alpha_{+}=\alpha_{+}(\Delta,z,\bar{z})$
corresponds to the analogous quantity for positive helicity.

Thus, we have recovered the superfields $\Psi$ (for $\mathcal{N}=8$
Supergravity) and $Z$ (for $\mathcal{N}=4$ SYM theory). Consequently,
the holomorphic potential density takes the form:
\begin{equation}
h_{\sigma}^{a}\left(z,\bar{z}\big|\mathbb{X}\right)\coloneqq\int_{\mathcal{P}}\frac{d\Delta}{2\pi i}\tilde{\phi}_{\Delta}(z,\bar{z}\big|\mathbb{X})W_{\sigma,\Delta}(z,\bar{z})\mathsf{S}^{a}.
\end{equation}

\paragraph{Scaling Reduction.}

We now proceed to motivate the scaling reduction from twistor to minitwistor
space. It is pertinent to recall that our application of twistor theory
serves primarily as a heuristic guide. 

The expression for $h_{\sigma}^{a}$ can be reformulated as an integral
over a two-component spinor, as follows:
\begin{equation}
h_{\sigma}^{a}\left(z,\bar{z}\big|\mathbb{X}\right)\coloneqq\int_{\mathcal{P}}\frac{d\Delta}{2\pi i}\int d^{2}\mu\,\frac{\Gamma\left(\Delta\right)}{\left(-i\mu\cdot\bar{\pi}+\xi^{\alpha}\bar{\xi}_{\alpha}\right)}f_{\sigma,\Delta}^{a},
\end{equation}
where $f_{\sigma,\Delta}^{a}$ is defined on the holomorphically extended
twistor space as:
\begin{equation}
f_{\sigma,\Delta}^{a}(z,\bar{z}\big|X,\mu_{\dot{A}},\pi^{A})\coloneqq\delta^{\left(2\right)}(\mu_{\dot{A}}-\pi^{A}X_{A\dot{A}})W_{\sigma,\Delta}(z,\bar{z})\mathsf{S}^{a}.
\end{equation}
Here, the equality $\mu_{\dot{A}}-\pi^{A}X_{A\dot{A}}=0$ corresponds
to the twistor equation, as explained in Appendix A of \citet{witten2004perturbative}.
From Sections II and III, we have learned that the multi-graviton
and multi-gluon celestial amplitudes in a MHV configuration admit
a natural geometric interpretation within the minitwistor space $\mathbf{MT}$
of Euclidean $AdS_{3}$.

Following \citet{bu2023celestial}, the scaling reduction is now performed
by identifying two-component spinors $\mu_{\dot{A}}\simeq\mu'_{\dot{A}}$
whenever there exists a \emph{complex }scalar $a\in\mathbf{C}$ such
that $\mu_{\dot{A}}=a\mu'_{\dot{A}}$. Under this equivalence, the
scale-reduced counterpart of $f_{\sigma,\Delta}^{a}$ is given by:
\begin{equation}
\tilde{f}_{\sigma,\Delta}^{a}(z,\bar{z}\big|x^{\mu},\mu_{\dot{A}},\pi^{A})\coloneqq\delta^{\left(2\right)}(\mu_{\dot{A}}-\pi^{A}x_{A\dot{A}})W_{\sigma,\Delta}(z,\bar{z})\mathsf{S}^{a},
\end{equation}
where $x_{A\dot{A}}\in\mathbf{H}_{3}$ is a point on the hyperboloid
and $(\mu_{\dot{A}},\pi^{A})\in\mathbf{MT}$ is a point on the minitwistor
space of $\mathbf{H}_{3}$. 

Accordingly, the scale-reduced holomorphic potential density takes
the form:
\begin{equation}
\tilde{h}_{\sigma,\Delta}^{a}(z,\bar{z}\big|x^{\mu},\mu_{\dot{A}},\pi^{A})=\int_{\mathcal{P}}\frac{d\Delta}{2\pi i}\tilde{G}_{\Delta}(z,\bar{z}\big|\mathbb{X})W_{\sigma,\Delta}(z,\bar{z})\mathsf{S}^{a},\label{eq:Holomorphic-Potential-Density}
\end{equation}
where $\tilde{G}_{\Delta}$ denotes the chiral supersymmetric extension
of the bulk-to-boundary Green's function on the $\mathcal{N}$-supersymmetric
hyperboloid $\mathbf{H}_{3}$. Thus, we recovered the lifted potentials
$\boldsymbol{\Omega}$ for $\mathcal{N}=8$ Supergravity, introduced
in Section II, and $\mathscr{A}$ for $\mathcal{N}=4$ SYM theory,
introduced in Section III, as we wished.
\begin{rem*}
From a mathematical perspective, the geometric structure underlying
our holographic models is that of a holomorphic line bundle, wherein
the complex projective line $\mathbf{CP}^{1}$ serves as the fibre
over the $\mathcal{N}$-extended supersymmetric $AdS_{3}$. The superpotential
$\tilde{h}_{\sigma,\Delta}^{a}$ is naturally identified as a section
of this fibre bundle.
\end{rem*}

\subsection{$AdS_{3}$ Action Integral\label{subsec:Superspace-Effective-Action}}

In this concluding section, we propose the action integral governing
the holographic $AdS_{3}$ models, which reproduces the tree-level
MHV celestial leaf super-amplitudes. Let $\mathcal{N}$ denote the
number of supersymmetry generators, and let $\mathcal{G}$ represent
the gauge group associated with special orthogonal matrices. For notational
simplicity, we define:
\begin{equation}
\int d\mathbb{X}\,(\cdot)\coloneqq\int_{\mathbf{H}_{3}}d^{3}x\int d^{\mathcal{N}}\theta\,(\cdot).
\end{equation}

Let $\mathcal{S}[G]$ denote the action functional for the level-one
$\mathcal{G}$ WZNW model on $\mathbf{CP}^{1}$. We postulate the
following action for the holographic $AdS_{3}$ models: 
\begin{equation}
\mathcal{I}\left[G,H,e\right]\coloneqq-\int_{\mathbf{H}_{3}}d\mathbb{X}\,\mathcal{S}\left[G\right]+\Theta\left[G,H,e\right],\label{eq:Effective-Action}
\end{equation}
where the term $\Theta\left[G,H,e\right]$ is defined by:
\begin{equation}
\Theta\left[G,H,e\right]\coloneqq\frac{1}{\pi}\int d\mathbb{X}\int d^{2}z\,\left(\text{Tr}\left[H_{z}\left(\partial_{\bar{z}}G\right)G^{-1}\right]+e(H_{\bar{z}}+\left(\partial_{\bar{z}}G\right)G^{-1})\right).\label{eq:Theta}
\end{equation}
In this expression, $\text{Tr}(\cdot)$ denotes the trace taken over
the matrix Lie group $\mathcal{G}$, and $e$ is a Lagrange multiplier.

\subsubsection{Euler-Lagrange Equations}

The variation of Eq. (\ref{eq:Effective-Action}) with respect to
the Lagrange multiplier $e$ trivially yields the constraint equation:
\begin{equation}
H_{\bar{z}}+(\partial_{\bar{z}}G)G^{-1}=0.
\end{equation}

To compute the variation of the WZNW action $\mathcal{S}[G]$, we
apply the Polyakov-Wiegmann identity, originally derived in \citet{polyakov1983theory}.
For two $\mathcal{G}$-valued matrix fields $M=M\left(z,\bar{z}\right)$
and $N=N\left(z,\bar{z}\right)$, the identity reads:
\begin{equation}
\mathcal{S}[MN]=\mathcal{S}[M]+\mathcal{S}[N]-\frac{1}{\pi}\int d^{2}z\,[\text{Tr}M^{-1}(\partial_{\bar{z}}M)(\partial_{z}N)N^{-1}].
\end{equation}
Let $\left\Vert A\right\Vert \coloneqq(\text{Tr}A^{T}A)^{1/2}$ define
the norm on the matrix Lie group $\mathcal{G}$, and let $\lambda$
denote an ``infinitesimal'' matrix field. Using the Polyakov-Wiegmann
identity, we compute the variation:
\begin{equation}
\mathcal{S}[G^{T}(1+\lambda)]-\mathcal{S}[G^{T}]=\frac{1}{\pi}\int d^{2}z\,\text{Tr}[\partial_{z}(G^{T-1}(\partial_{\bar{z}}G^{T})\lambda)]+\mathcal{O}(\left\Vert \lambda\right\Vert ^{2}).
\end{equation}
By setting $\lambda=G^{T-1}\delta G^{T}$, the variation of $\mathcal{S}$
with respect to $G^{T}$ becomes:
\begin{equation}
\delta_{G^{T}}\mathcal{S}=\frac{1}{\pi}\int d^{2}z\,\text{Tr}[\partial_{z}(G^{T-1}(\partial_{\bar{z}}G^{T}))G^{T-1}\delta G^{T}]+\mathcal{O}(\left\Vert \lambda\right\Vert ^{2}).
\end{equation}

The next step is to evaluate the variation of the term $\Theta$ with
respect to $G^{T}$, under the constraint $H_{\bar{z}}+(\partial_{\bar{z}}G)G^{-1}=0$.
Rewriting $\Theta$ in terms of $G^{T}$, we have:
\begin{equation}
\left(\Theta[G,H,e]\right){}_{\frac{\delta\mathcal{I}}{\delta e}=0}=-\frac{1}{\pi}\int d\mathbb{X}\int d^{2}z\,\text{Tr}(H_{z}G^{T-1}(\partial_{\bar{z}}G^{T})).
\end{equation}
The variation $\delta_{G^{T}}\Theta$ is the computed as:
\begin{equation}
\delta_{G^{T}}\Theta=\frac{1}{\pi}\int d\mathbb{X}\int d^{2}z\,\text{Tr}[(G^{T-1}(\partial_{\bar{z}}G^{T})H_{z}G^{T-1}+\partial_{\bar{z}}(H_{z}G^{T-1}))\delta G^{T}].
\end{equation}
Therefore, the equation of motion obtained from the variation of $\mathcal{I}$
with respect to $G^{T}$ becomes:
\begin{equation}
-\partial_{z}((\partial_{\bar{z}}G)G^{-1})=-(\partial_{\bar{z}}G)G^{-1}H_{z}+\partial_{\bar{z}}H_{z}+H_{z}(\partial_{\bar{z}}G)G^{-1}.
\end{equation}
Substituting the constraint $H_{\bar{z}}+(\partial_{\bar{z}}G)G^{-1}=0$,
we find:
\begin{equation}
\partial_{z}H_{\bar{z}}=\partial_{\bar{z}}H_{z}+[H_{\bar{z}},H_{z}]=0,
\end{equation}
which precisely corresponds to the ``zero-curvature'' condition identified
by \citet{abe2005multigluon} and \citet{abe5analysis} as the residual
superspace constraint equations in the chiral semi-analytic gauge.

\subsubsection{The On-shell Effective Action}

We now demonstrate that the on-shell effective action $\mathcal{W}_{AdS_{3}}$
derived from $\mathcal{I}$ is the generating function for the leaf
amplitudes.

To begin, consider a $\mathcal{G}$-valued matrix field $M=M(z,\bar{z})$
satisfying $H_{z}=M^{T}\partial_{z}M^{T-1}$. Substituting this decomposition
of the holomorphic superpotential into the equation of motion obtained
above, we deduce that $M=G^{T}$. Thus, the solution $H=H_{z}dz+H_{\bar{z}}d\bar{z}$
to the equations of motion, which is equivalently a solution to the
superspace constraints in the chiral semi-analytic gauge, is entirely
characterised by $G$, with
\begin{equation}
H_{z}=-(\partial_{z}G)G^{-1},\,\,\,H_{\bar{z}}=-(\partial_{\bar{z}}G)G^{-1}.
\end{equation}
Consequently, the functional $\Theta$, evaluated on-shell with $\delta\mathcal{I}/\delta G=\delta\mathcal{I}/\delta e=0$,
becomes:
\begin{equation}
\left(\Theta[G,H,e]\right){}_{\frac{\delta\mathcal{I}}{\delta G^{T}}=\frac{\delta\mathcal{I}}{\delta e}=0}=-\frac{1}{\pi}\int d\mathbb{X}\int d^{2}z\,\text{Tr}[(\partial_{z}G)G^{-1}(\partial_{\bar{z}}G)G^{-1}].
\end{equation}

To proceed, we employ the Polyakov-Wiegmann formula to establish the
following identity:
\begin{equation}
\mathcal{S}[G^{T}G]=2\mathcal{S}[G]-\frac{1}{\pi}\int d^{2}z\,\text{Tr}[G^{T-1}(\partial_{\bar{z}}G^{T})(\partial_{z}G)G^{-1}]=0.
\end{equation}
From this result, we deduce that:
\begin{equation}
\left(\Theta[G,H,e]\right){}_{\frac{\delta\mathcal{I}}{\delta G}=\frac{\delta\mathcal{I}}{\delta e}=0}=2\int d\mathbb{X}\,\mathcal{S}[G].
\end{equation}
Substituting this expression into Eq. (\ref{eq:Effective-Action}),
we obtain:
\begin{equation}
\mathcal{W}_{AdS_{3}}\coloneqq(\mathcal{I}[G,H,e])_{\frac{\delta\mathcal{I}}{\delta G}=\frac{\delta\mathcal{I}}{\delta e}=0}=\int d\mathbb{X}\,\mathcal{S}[G].
\end{equation}

Finally, by substituting the holomorphic potential density, $h_{\sigma,\Delta}^{a}$,
as derived in Eq. (\ref{eq:Holomorphic-Potential-Density}), into
the on-shell effective action, we recover the generating functional
for the leaf amplitudes, thereby completing the proof.

\bibliographystyle{../TTCFT/revtex-tds/bibtex/bst/revtex/aipnum4-2}
\bibliography{CCFT2}

\begin{thebibliography}{52}%
\makeatletter
\providecommand \@ifxundefined [1]{%
 \@ifx{#1\undefined}
}%
\providecommand \@ifnum [1]{%
 \ifnum #1\expandafter \@firstoftwo
 \else \expandafter \@secondoftwo
 \fi
}%
\providecommand \@ifx [1]{%
 \ifx #1\expandafter \@firstoftwo
 \else \expandafter \@secondoftwo
 \fi
}%
\providecommand \natexlab [1]{#1}%
\providecommand \enquote  [1]{``#1''}%
\providecommand \bibnamefont  [1]{#1}%
\providecommand \bibfnamefont [1]{#1}%
\providecommand \citenamefont [1]{#1}%
\providecommand \href@noop [0]{\@secondoftwo}%
\providecommand \href [0]{\begingroup \@sanitize@url \@href}%
\providecommand \@href[1]{\@@startlink{#1}\@@href}%
\providecommand \@@href[1]{\endgroup#1\@@endlink}%
\providecommand \@sanitize@url [0]{\catcode `\\12\catcode `\$12\catcode
  `\&12\catcode `\#12\catcode `\^12\catcode `\_12\catcode `\%12\relax}%
\providecommand \@@startlink[1]{}%
\providecommand \@@endlink[0]{}%
\providecommand \url  [0]{\begingroup\@sanitize@url \@url }%
\providecommand \@url [1]{\endgroup\@href {#1}{\urlprefix }}%
\providecommand \urlprefix  [0]{URL }%
\providecommand \Eprint [0]{\href }%
\providecommand \doibase [0]{https://doi.org/}%
\providecommand \selectlanguage [0]{\@gobble}%
\providecommand \bibinfo  [0]{\@secondoftwo}%
\providecommand \bibfield  [0]{\@secondoftwo}%
\providecommand \translation [1]{[#1]}%
\providecommand \BibitemOpen [0]{}%
\providecommand \bibitemStop [0]{}%
\providecommand \bibitemNoStop [0]{.\EOS\space}%
\providecommand \EOS [0]{\spacefactor3000\relax}%
\providecommand \BibitemShut  [1]{\csname bibitem#1\endcsname}%
\let\auto@bib@innerbib\@empty
\bibitem [{\citenamefont {Ogawa}\ \emph {et~al.}(2024)\citenamefont {Ogawa},
  \citenamefont {Takahashi}, \citenamefont {Tsuda},\ and\ \citenamefont
  {Waki}}]{ogawa2024celestial}%
  \BibitemOpen
  \bibfield  {author} {\bibinfo {author} {\bibfnamefont {N.}~\bibnamefont
  {Ogawa}}, \bibinfo {author} {\bibfnamefont {S.}~\bibnamefont {Takahashi}},
  \bibinfo {author} {\bibfnamefont {T.}~\bibnamefont {Tsuda}},\ and\ \bibinfo
  {author} {\bibfnamefont {T.}~\bibnamefont {Waki}},\ }\href@noop {} {\bibfield
   {journal} {\bibinfo  {journal} {arXiv preprint arXiv:2404.12049}\ }
  (\bibinfo {year} {2024})}\BibitemShut {NoStop}%
\bibitem [{\citenamefont {Mol}(2024{\natexlab{a}})}]{mol2024comments}%
  \BibitemOpen
  \bibfield  {author} {\bibinfo {author} {\bibfnamefont {I.}~\bibnamefont
  {Mol}},\ }\href@noop {} {\bibfield  {journal} {\bibinfo  {journal} {arXiv
  preprint arXiv:2410.02620}\ } (\bibinfo {year}
  {2024}{\natexlab{a}})}\BibitemShut {NoStop}%
\bibitem [{\citenamefont {Cachazo}\ and\ \citenamefont
  {Skinner}(2013)}]{cachazo2013gravity}%
  \BibitemOpen
  \bibfield  {author} {\bibinfo {author} {\bibfnamefont {F.}~\bibnamefont
  {Cachazo}}\ and\ \bibinfo {author} {\bibfnamefont {D.}~\bibnamefont
  {Skinner}},\ }\href@noop {} {\bibfield  {journal} {\bibinfo  {journal}
  {Physical Review Letters}\ }\textbf {\bibinfo {volume} {110}},\ \bibinfo
  {pages} {161301} (\bibinfo {year} {2013})}\BibitemShut {NoStop}%
\bibitem [{\citenamefont {Cachazo}\ \emph {et~al.}(2014)\citenamefont
  {Cachazo}, \citenamefont {Mason}, \citenamefont {Skinner} \emph
  {et~al.}}]{cachazo2014gravity}%
  \BibitemOpen
  \bibfield  {author} {\bibinfo {author} {\bibfnamefont {F.}~\bibnamefont
  {Cachazo}}, \bibinfo {author} {\bibfnamefont {L.}~\bibnamefont {Mason}},
  \bibinfo {author} {\bibfnamefont {D.}~\bibnamefont {Skinner}}, \emph
  {et~al.},\ }\href@noop {} {\bibfield  {journal} {\bibinfo  {journal} {SIGMA.
  Symmetry, Integrability and Geometry: Methods and Applications}\ }\textbf
  {\bibinfo {volume} {10}},\ \bibinfo {pages} {051} (\bibinfo {year}
  {2014})}\BibitemShut {NoStop}%
\bibitem [{\citenamefont {Adamo}\ and\ \citenamefont
  {Mason}(2013)}]{adamo2013twistor}%
  \BibitemOpen
  \bibfield  {author} {\bibinfo {author} {\bibfnamefont {T.}~\bibnamefont
  {Adamo}}\ and\ \bibinfo {author} {\bibfnamefont {L.}~\bibnamefont {Mason}},\
  }\href@noop {} {\bibfield  {journal} {\bibinfo  {journal} {Classical and
  Quantum Gravity}\ }\textbf {\bibinfo {volume} {30}},\ \bibinfo {pages}
  {075020} (\bibinfo {year} {2013})}\BibitemShut {NoStop}%
\bibitem [{\citenamefont {de~Boer}\ and\ \citenamefont
  {Solodukhin}(2003)}]{de2003holographic}%
  \BibitemOpen
  \bibfield  {author} {\bibinfo {author} {\bibfnamefont {J.}~\bibnamefont
  {de~Boer}}\ and\ \bibinfo {author} {\bibfnamefont {S.~N.}\ \bibnamefont
  {Solodukhin}},\ }\href@noop {} {\bibfield  {journal} {\bibinfo  {journal}
  {Nuclear Physics B}\ }\textbf {\bibinfo {volume} {665}},\ \bibinfo {pages}
  {545} (\bibinfo {year} {2003})}\BibitemShut {NoStop}%
\bibitem [{\citenamefont {Casali}, \citenamefont {Melton},\ and\ \citenamefont
  {Strominger}(2022)}]{casali2022celestial}%
  \BibitemOpen
  \bibfield  {author} {\bibinfo {author} {\bibfnamefont {E.}~\bibnamefont
  {Casali}}, \bibinfo {author} {\bibfnamefont {W.}~\bibnamefont {Melton}},\
  and\ \bibinfo {author} {\bibfnamefont {A.}~\bibnamefont {Strominger}},\
  }\href@noop {} {\bibfield  {journal} {\bibinfo  {journal} {Journal of High
  Energy Physics}\ }\textbf {\bibinfo {volume} {2022}},\ \bibinfo {pages} {1}
  (\bibinfo {year} {2022})}\BibitemShut {NoStop}%
\bibitem [{\citenamefont {Abe}, \citenamefont {Nair},\ and\ \citenamefont
  {Park}(2005)}]{abe2005multigluon}%
  \BibitemOpen
  \bibfield  {author} {\bibinfo {author} {\bibfnamefont {Y.}~\bibnamefont
  {Abe}}, \bibinfo {author} {\bibfnamefont {V.}~\bibnamefont {Nair}},\ and\
  \bibinfo {author} {\bibfnamefont {M.-I.}\ \bibnamefont {Park}},\ }\href@noop
  {} {\bibfield  {journal} {\bibinfo  {journal} {Physical Review D?Particles,
  Fields, Gravitation, and Cosmology}\ }\textbf {\bibinfo {volume} {71}},\
  \bibinfo {pages} {025002} (\bibinfo {year} {2005})}\BibitemShut {NoStop}%
\bibitem [{\citenamefont {Abe}()}]{abe5analysis}%
  \BibitemOpen
  \bibfield  {author} {\bibinfo {author} {\bibfnamefont {Y.}~\bibnamefont
  {Abe}},\ }\href@noop {} {\bibinfo  {journal} {SQS?05}\ ,\ \bibinfo {pages}
  {376}}\BibitemShut {NoStop}%
\bibitem [{\citenamefont {Galperin}\ \emph {et~al.}(2001)\citenamefont
  {Galperin}, \citenamefont {Ivanov}, \citenamefont {Ogievetsky},\ and\
  \citenamefont {Sokatchev}}]{galperin2001harmonic}%
  \BibitemOpen
\bibfield  {journal} {  }\bibfield  {author} {\bibinfo {author} {\bibfnamefont
  {A.~S.}\ \bibnamefont {Galperin}}, \bibinfo {author} {\bibfnamefont {E.~A.}\
  \bibnamefont {Ivanov}}, \bibinfo {author} {\bibfnamefont {V.~I.}\
  \bibnamefont {Ogievetsky}},\ and\ \bibinfo {author} {\bibfnamefont {E.~S.}\
  \bibnamefont {Sokatchev}},\ }\href@noop {} {\emph {\bibinfo {title} {Harmonic
  superspace}}}\ (\bibinfo  {publisher} {Cambridge University Press},\ \bibinfo
  {year} {2001})\BibitemShut {NoStop}%
\bibitem [{\citenamefont {Bu}\ and\ \citenamefont
  {Seet}(2023)}]{bu2023celestial}%
  \BibitemOpen
  \bibfield  {author} {\bibinfo {author} {\bibfnamefont {W.}~\bibnamefont
  {Bu}}\ and\ \bibinfo {author} {\bibfnamefont {S.}~\bibnamefont {Seet}},\
  }\href@noop {} {\bibfield  {journal} {\bibinfo  {journal} {Journal of High
  Energy Physics}\ }\textbf {\bibinfo {volume} {2023}},\ \bibinfo {pages} {1}
  (\bibinfo {year} {2023})}\BibitemShut {NoStop}%
\bibitem [{\citenamefont {Weinberg}(2010)}]{weinberg2010six}%
  \BibitemOpen
  \bibfield  {author} {\bibinfo {author} {\bibfnamefont {S.}~\bibnamefont
  {Weinberg}},\ }\href@noop {} {\bibfield  {journal} {\bibinfo  {journal}
  {Physical Review D?Particles, Fields, Gravitation, and Cosmology}\ }\textbf
  {\bibinfo {volume} {82}},\ \bibinfo {pages} {045031} (\bibinfo {year}
  {2010})}\BibitemShut {NoStop}%
\bibitem [{\citenamefont {DeWitt}(1992)}]{dewitt1992supermanifolds}%
  \BibitemOpen
  \bibfield  {author} {\bibinfo {author} {\bibfnamefont {B.~S.}\ \bibnamefont
  {DeWitt}},\ }\href@noop {} {\emph {\bibinfo {title} {Supermanifolds}}}\
  (\bibinfo  {publisher} {Cambridge University Press},\ \bibinfo {year}
  {1992})\BibitemShut {NoStop}%
\bibitem [{\citenamefont {Rogers}(2007)}]{rogers2007supermanifolds}%
  \BibitemOpen
  \bibfield  {author} {\bibinfo {author} {\bibfnamefont {A.}~\bibnamefont
  {Rogers}},\ }\href@noop {} {\emph {\bibinfo {title} {Supermanifolds: theory
  and applications}}}\ (\bibinfo  {publisher} {World Scientific},\ \bibinfo
  {year} {2007})\BibitemShut {NoStop}%
\bibitem [{\citenamefont {Berezin}(2013)}]{berezin2013introduction}%
  \BibitemOpen
  \bibfield  {author} {\bibinfo {author} {\bibfnamefont {F.~A.}\ \bibnamefont
  {Berezin}},\ }\href@noop {} {\emph {\bibinfo {title} {Introduction to
  superanalysis}}},\ Vol.~\bibinfo {volume} {9}\ (\bibinfo  {publisher}
  {Springer Science \& Business Media},\ \bibinfo {year} {2013})\BibitemShut
  {NoStop}%
\bibitem [{\citenamefont {Leites}(1980)}]{leites1980introduction}%
  \BibitemOpen
  \bibfield  {author} {\bibinfo {author} {\bibfnamefont {D.~A.}\ \bibnamefont
  {Leites}},\ }\href@noop {} {\bibfield  {journal} {\bibinfo  {journal}
  {Russian Mathematical Surveys}\ }\textbf {\bibinfo {volume} {35}},\ \bibinfo
  {pages} {1} (\bibinfo {year} {1980})}\BibitemShut {NoStop}%
\bibitem [{\citenamefont {Batchelor}(1979)}]{batchelor1979structure}%
  \BibitemOpen
  \bibfield  {author} {\bibinfo {author} {\bibfnamefont {M.}~\bibnamefont
  {Batchelor}},\ }\href@noop {} {\bibfield  {journal} {\bibinfo  {journal}
  {Transactions of the American Mathematical Society}\ }\textbf {\bibinfo
  {volume} {253}},\ \bibinfo {pages} {329} (\bibinfo {year}
  {1979})}\BibitemShut {NoStop}%
\bibitem [{\citenamefont {Batchelor}(1980)}]{batchelor1980two}%
  \BibitemOpen
  \bibfield  {author} {\bibinfo {author} {\bibfnamefont {M.}~\bibnamefont
  {Batchelor}},\ }\href@noop {} {\bibfield  {journal} {\bibinfo  {journal}
  {Transactions of the American Mathematical Society}\ }\textbf {\bibinfo
  {volume} {258}},\ \bibinfo {pages} {257} (\bibinfo {year}
  {1980})}\BibitemShut {NoStop}%
\bibitem [{\citenamefont {Mol}(2024{\natexlab{b}})}]{mol2024holographic}%
  \BibitemOpen
  \bibfield  {author} {\bibinfo {author} {\bibfnamefont {I.}~\bibnamefont
  {Mol}},\ }\href@noop {} {\bibfield  {journal} {\bibinfo  {journal} {arXiv
  preprint arXiv:2408.10944}\ } (\bibinfo {year}
  {2024}{\natexlab{b}})}\BibitemShut {NoStop}%
\bibitem [{\citenamefont {Berends}, \citenamefont {Giele},\ and\ \citenamefont
  {Kuijf}(1988)}]{berends1988relations}%
  \BibitemOpen
  \bibfield  {author} {\bibinfo {author} {\bibfnamefont {F.~A.}\ \bibnamefont
  {Berends}}, \bibinfo {author} {\bibfnamefont {W.}~\bibnamefont {Giele}},\
  and\ \bibinfo {author} {\bibfnamefont {H.}~\bibnamefont {Kuijf}},\
  }\href@noop {} {\bibfield  {journal} {\bibinfo  {journal} {Physics Letters
  B}\ }\textbf {\bibinfo {volume} {211}},\ \bibinfo {pages} {91} (\bibinfo
  {year} {1988})}\BibitemShut {NoStop}%
\bibitem [{\citenamefont {Nair}(1988)}]{nair1988current}%
  \BibitemOpen
  \bibfield  {author} {\bibinfo {author} {\bibfnamefont {V.~P.}\ \bibnamefont
  {Nair}},\ }\href@noop {} {\bibfield  {journal} {\bibinfo  {journal} {Physics
  Letters B}\ }\textbf {\bibinfo {volume} {214}},\ \bibinfo {pages} {215}
  (\bibinfo {year} {1988})}\BibitemShut {NoStop}%
\bibitem [{\citenamefont {Nair}(2005{\natexlab{a}})}]{nair2005note}%
  \BibitemOpen
  \bibfield  {author} {\bibinfo {author} {\bibfnamefont {V.}~\bibnamefont
  {Nair}},\ }\href@noop {} {\bibfield  {journal} {\bibinfo  {journal} {Physical
  Review D?Particles, Fields, Gravitation, and Cosmology}\ }\textbf {\bibinfo
  {volume} {71}},\ \bibinfo {pages} {121701} (\bibinfo {year}
  {2005}{\natexlab{a}})}\BibitemShut {NoStop}%
\bibitem [{\citenamefont {Pasterski}\ and\ \citenamefont
  {Shao}(2017)}]{pasterski2017conformal}%
  \BibitemOpen
  \bibfield  {author} {\bibinfo {author} {\bibfnamefont {S.}~\bibnamefont
  {Pasterski}}\ and\ \bibinfo {author} {\bibfnamefont {S.-H.}\ \bibnamefont
  {Shao}},\ }\href@noop {} {\bibfield  {journal} {\bibinfo  {journal} {Physical
  Review D}\ }\textbf {\bibinfo {volume} {96}},\ \bibinfo {pages} {065022}
  (\bibinfo {year} {2017})}\BibitemShut {NoStop}%
\bibitem [{\citenamefont {Melton}, \citenamefont {Sharma},\ and\ \citenamefont
  {Strominger}(2023)}]{melton2023celestial}%
  \BibitemOpen
  \bibfield  {author} {\bibinfo {author} {\bibfnamefont {W.}~\bibnamefont
  {Melton}}, \bibinfo {author} {\bibfnamefont {A.}~\bibnamefont {Sharma}},\
  and\ \bibinfo {author} {\bibfnamefont {A.}~\bibnamefont {Strominger}},\
  }\href@noop {} {\bibfield  {journal} {\bibinfo  {journal} {arXiv preprint
  arXiv:2312.07820}\ } (\bibinfo {year} {2023})}\BibitemShut {NoStop}%
\bibitem [{\citenamefont {Melton}\ \emph {et~al.}(2024)\citenamefont {Melton},
  \citenamefont {Sharma}, \citenamefont {Strominger},\ and\ \citenamefont
  {Wang}}]{melton2024celestial}%
  \BibitemOpen
  \bibfield  {author} {\bibinfo {author} {\bibfnamefont {W.}~\bibnamefont
  {Melton}}, \bibinfo {author} {\bibfnamefont {A.}~\bibnamefont {Sharma}},
  \bibinfo {author} {\bibfnamefont {A.}~\bibnamefont {Strominger}},\ and\
  \bibinfo {author} {\bibfnamefont {T.}~\bibnamefont {Wang}},\ }\href@noop {}
  {\bibfield  {journal} {\bibinfo  {journal} {arXiv preprint arXiv:2403.18896}\
  } (\bibinfo {year} {2024})}\BibitemShut {NoStop}%
\bibitem [{\citenamefont {Teschner}(1999)}]{teschner1999mini}%
  \BibitemOpen
  \bibfield  {author} {\bibinfo {author} {\bibfnamefont {J.}~\bibnamefont
  {Teschner}},\ }\href@noop {} {\bibfield  {journal} {\bibinfo  {journal}
  {Nuclear Physics B}\ }\textbf {\bibinfo {volume} {546}},\ \bibinfo {pages}
  {369} (\bibinfo {year} {1999})}\BibitemShut {NoStop}%
\bibitem [{\citenamefont {Penedones}(2017)}]{penedones2017tasi}%
  \BibitemOpen
  \bibfield  {author} {\bibinfo {author} {\bibfnamefont {J.}~\bibnamefont
  {Penedones}},\ }in\ \href@noop {} {\emph {\bibinfo {booktitle} {New Frontiers
  in Fields and Strings: TASI 2015 Proceedings of the 2015 Theoretical Advanced
  Study Institute in Elementary Particle Physics}}}\ (\bibinfo  {publisher}
  {World Scientific},\ \bibinfo {year} {2017})\ pp.\ \bibinfo {pages}
  {75--136}\BibitemShut {NoStop}%
\bibitem [{\citenamefont {Costa}, \citenamefont {Gon{\c{c}}alves},\ and\
  \citenamefont {Penedones}(2014)}]{costa2014spinning}%
  \BibitemOpen
  \bibfield  {author} {\bibinfo {author} {\bibfnamefont {M.~S.}\ \bibnamefont
  {Costa}}, \bibinfo {author} {\bibfnamefont {V.}~\bibnamefont
  {Gon{\c{c}}alves}},\ and\ \bibinfo {author} {\bibfnamefont {J.}~\bibnamefont
  {Penedones}},\ }\href@noop {} {\bibfield  {journal} {\bibinfo  {journal}
  {Journal of High Energy Physics}\ }\textbf {\bibinfo {volume} {2014}},\
  \bibinfo {pages} {1} (\bibinfo {year} {2014})}\BibitemShut {NoStop}%
\bibitem [{\citenamefont {Ribault}\ and\ \citenamefont
  {Teschner}(2005)}]{ribault2005h3+}%
  \BibitemOpen
  \bibfield  {author} {\bibinfo {author} {\bibfnamefont {S.}~\bibnamefont
  {Ribault}}\ and\ \bibinfo {author} {\bibfnamefont {J.}~\bibnamefont
  {Teschner}},\ }\href@noop {} {\bibfield  {journal} {\bibinfo  {journal}
  {Journal of High Energy Physics}\ }\textbf {\bibinfo {volume} {2005}},\
  \bibinfo {pages} {014} (\bibinfo {year} {2005})}\BibitemShut {NoStop}%
\bibitem [{\citenamefont {Giveon}, \citenamefont {Kutasov},\ and\ \citenamefont
  {Seiberg}(1998)}]{giveon1998comments}%
  \BibitemOpen
  \bibfield  {author} {\bibinfo {author} {\bibfnamefont {A.}~\bibnamefont
  {Giveon}}, \bibinfo {author} {\bibfnamefont {D.}~\bibnamefont {Kutasov}},\
  and\ \bibinfo {author} {\bibfnamefont {N.}~\bibnamefont {Seiberg}},\
  }\href@noop {} {\bibfield  {journal} {\bibinfo  {journal} {arXiv preprint
  hep-th/9806194}\ } (\bibinfo {year} {1998})}\BibitemShut {NoStop}%
\bibitem [{\citenamefont {Dolan}\ and\ \citenamefont
  {Witten}(1999)}]{dolan1999vertex}%
  \BibitemOpen
  \bibfield  {author} {\bibinfo {author} {\bibfnamefont {L.}~\bibnamefont
  {Dolan}}\ and\ \bibinfo {author} {\bibfnamefont {E.}~\bibnamefont {Witten}},\
  }\href@noop {} {\bibfield  {journal} {\bibinfo  {journal} {Journal of High
  Energy Physics}\ }\textbf {\bibinfo {volume} {1999}},\ \bibinfo {pages} {003}
  (\bibinfo {year} {1999})}\BibitemShut {NoStop}%
\bibitem [{\citenamefont {Dolan}(2001)}]{dolan2001vertex}%
  \BibitemOpen
  \bibfield  {author} {\bibinfo {author} {\bibfnamefont {L.}~\bibnamefont
  {Dolan}},\ }\href@noop {} {\bibfield  {journal} {\bibinfo  {journal}
  {International Journal of Modern Physics A}\ }\textbf {\bibinfo {volume}
  {16}},\ \bibinfo {pages} {812} (\bibinfo {year} {2001})}\BibitemShut
  {NoStop}%
\bibitem [{\citenamefont {Nair}(2005{\natexlab{b}})}]{nair2005chern}%
  \BibitemOpen
  \bibfield  {author} {\bibinfo {author} {\bibfnamefont {V.}~\bibnamefont
  {Nair}},\ }\href@noop {} {\bibfield  {journal} {\bibinfo  {journal} {Notes
  for lectures at BUSSTEPP}\ } (\bibinfo {year}
  {2005}{\natexlab{b}})}\BibitemShut {NoStop}%
\bibitem [{\citenamefont {Parke}\ and\ \citenamefont
  {Taylor}(1986)}]{parke1986amplitude}%
  \BibitemOpen
  \bibfield  {author} {\bibinfo {author} {\bibfnamefont {S.~J.}\ \bibnamefont
  {Parke}}\ and\ \bibinfo {author} {\bibfnamefont {T.~R.}\ \bibnamefont
  {Taylor}},\ }\href@noop {} {\bibfield  {journal} {\bibinfo  {journal}
  {Physical Review Letters}\ }\textbf {\bibinfo {volume} {56}},\ \bibinfo
  {pages} {2459} (\bibinfo {year} {1986})}\BibitemShut {NoStop}%
\bibitem [{\citenamefont {Berends}\ and\ \citenamefont
  {Giele}(1988)}]{berends1988recursive}%
  \BibitemOpen
  \bibfield  {author} {\bibinfo {author} {\bibfnamefont {F.~A.}\ \bibnamefont
  {Berends}}\ and\ \bibinfo {author} {\bibfnamefont {W.}~\bibnamefont
  {Giele}},\ }\href@noop {} {\bibfield  {journal} {\bibinfo  {journal} {Nuclear
  Physics B}\ }\textbf {\bibinfo {volume} {306}},\ \bibinfo {pages} {759}
  (\bibinfo {year} {1988})}\BibitemShut {NoStop}%
\bibitem [{\citenamefont {Elvang}\ and\ \citenamefont
  {Huang}(2013)}]{elvang2013scattering}%
  \BibitemOpen
  \bibfield  {author} {\bibinfo {author} {\bibfnamefont {H.}~\bibnamefont
  {Elvang}}\ and\ \bibinfo {author} {\bibfnamefont {Y.-t.}\ \bibnamefont
  {Huang}},\ }\href@noop {} {\bibfield  {journal} {\bibinfo  {journal} {arXiv
  preprint arXiv:1308.1697}\ } (\bibinfo {year} {2013})}\BibitemShut {NoStop}%
\bibitem [{\citenamefont {Badger}\ \emph {et~al.}(2024)\citenamefont {Badger},
  \citenamefont {Henn}, \citenamefont {Plefka},\ and\ \citenamefont
  {Zoia}}]{badger2024scattering}%
  \BibitemOpen
  \bibfield  {author} {\bibinfo {author} {\bibfnamefont {S.}~\bibnamefont
  {Badger}}, \bibinfo {author} {\bibfnamefont {J.}~\bibnamefont {Henn}},
  \bibinfo {author} {\bibfnamefont {J.~C.}\ \bibnamefont {Plefka}},\ and\
  \bibinfo {author} {\bibfnamefont {S.}~\bibnamefont {Zoia}},\ }\href@noop {}
  {\emph {\bibinfo {title} {Scattering Amplitudes in Quantum Field Theory}}}\
  (\bibinfo  {publisher} {Springer Nature},\ \bibinfo {year}
  {2024})\BibitemShut {NoStop}%
\bibitem [{\citenamefont {Witten}(2004)}]{witten2004perturbative}%
  \BibitemOpen
  \bibfield  {author} {\bibinfo {author} {\bibfnamefont {E.}~\bibnamefont
  {Witten}},\ }\href@noop {} {\bibfield  {journal} {\bibinfo  {journal}
  {Communications in Mathematical Physics}\ }\textbf {\bibinfo {volume}
  {252}},\ \bibinfo {pages} {189} (\bibinfo {year} {2004})}\BibitemShut
  {NoStop}%
\bibitem [{\citenamefont {Kostant}(2006)}]{kostant2006graded}%
  \BibitemOpen
  \bibfield  {author} {\bibinfo {author} {\bibfnamefont {B.}~\bibnamefont
  {Kostant}},\ }in\ \href@noop {} {\emph {\bibinfo {booktitle} {Differential
  Geometrical Methods in Mathematical Physics: Proceedings of the Symposium
  Held at the University of Bonn, July 1--4, 1975}}}\ (\bibinfo {organization}
  {Springer},\ \bibinfo {year} {2006})\ pp.\ \bibinfo {pages}
  {177--306}\BibitemShut {NoStop}%
\bibitem [{\citenamefont {Weinberg}(2005)}]{weinberg2005quantum}%
  \BibitemOpen
  \bibfield  {author} {\bibinfo {author} {\bibfnamefont {S.}~\bibnamefont
  {Weinberg}},\ }\href {https://books.google.com.br/books?id=jV19vAEACAAJ}
  {\emph {\bibinfo {title} {The Quantum Theory of Fields: Volume 2, Modern
  Applications}}}\ (\bibinfo  {publisher} {Cambridge University Press},\
  \bibinfo {year} {2005})\BibitemShut {NoStop}%
\bibitem [{\citenamefont {Lukierski}(1987)}]{Lukierski:1986kg}%
  \BibitemOpen
  \bibfield  {author} {\bibinfo {author} {\bibfnamefont {J.}~\bibnamefont
  {Lukierski}},\ }\href {https://doi.org/10.1007/3-540-17925-9_36} {\bibfield
  {journal} {\bibinfo  {journal} {Lect. Notes Phys.}\ }\textbf {\bibinfo
  {volume} {280}},\ \bibinfo {pages} {137} (\bibinfo {year}
  {1987})}\BibitemShut {NoStop}%
\bibitem [{\citenamefont {Lukierski}\ and\ \citenamefont
  {Nowicki}(1988)}]{Lukierski:1988vw}%
  \BibitemOpen
  \bibfield  {author} {\bibinfo {author} {\bibfnamefont {J.}~\bibnamefont
  {Lukierski}}\ and\ \bibinfo {author} {\bibfnamefont {A.}~\bibnamefont
  {Nowicki}},\ }\href {https://doi.org/10.1016/0370-2693(88)90903-3} {\bibfield
   {journal} {\bibinfo  {journal} {Phys. Lett. B}\ }\textbf {\bibinfo {volume}
  {211}},\ \bibinfo {pages} {276} (\bibinfo {year} {1988})}\BibitemShut
  {NoStop}%
\bibitem [{\citenamefont {Kotrla}\ and\ \citenamefont
  {Niederle}(1985)}]{Kotrla:1984ky}%
  \BibitemOpen
  \bibfield  {author} {\bibinfo {author} {\bibfnamefont {M.}~\bibnamefont
  {Kotrla}}\ and\ \bibinfo {author} {\bibfnamefont {J.}~\bibnamefont
  {Niederle}},\ }\href {https://doi.org/10.1007/BF01595531} {\bibfield
  {journal} {\bibinfo  {journal} {Czech. J. Phys. B}\ }\textbf {\bibinfo
  {volume} {35}},\ \bibinfo {pages} {602} (\bibinfo {year} {1985})}\BibitemShut
  {NoStop}%
\bibitem [{\citenamefont {Popov}\ and\ \citenamefont
  {Saemann}(2005)}]{Popov:2004rb}%
  \BibitemOpen
  \bibfield  {author} {\bibinfo {author} {\bibfnamefont {A.~D.}\ \bibnamefont
  {Popov}}\ and\ \bibinfo {author} {\bibfnamefont {C.}~\bibnamefont
  {Saemann}},\ }\href {https://doi.org/10.4310/ATMP.2005.v9.n6.a2} {\bibfield
  {journal} {\bibinfo  {journal} {Adv. Theor. Math. Phys.}\ }\textbf {\bibinfo
  {volume} {9}},\ \bibinfo {pages} {931} (\bibinfo {year} {2005})},\ \Eprint
  {https://arxiv.org/abs/hep-th/0405123} {arXiv:hep-th/0405123} \BibitemShut
  {NoStop}%
\bibitem [{\citenamefont {Wolf}(2006)}]{Wolf:2006me}%
  \BibitemOpen
  \bibfield  {author} {\bibinfo {author} {\bibfnamefont {M.}~\bibnamefont
  {Wolf}},\ }\emph {\bibinfo {title} {{On Supertwistor geometry and
  integrability in super gauge theory}}},\ \href@noop {} {\bibinfo {type}
  {Other thesis}} (\bibinfo {year} {2006}),\ \Eprint
  {https://arxiv.org/abs/hep-th/0611013} {arXiv:hep-th/0611013} \BibitemShut
  {NoStop}%
\bibitem [{\citenamefont {Ivanov}(2023)}]{ivanov2023n}%
  \BibitemOpen
  \bibfield  {author} {\bibinfo {author} {\bibfnamefont {E.}~\bibnamefont
  {Ivanov}},\ }in\ \href@noop {} {\emph {\bibinfo {booktitle} {Handbook of
  Quantum Gravity}}}\ (\bibinfo  {publisher} {Springer},\ \bibinfo {year}
  {2023})\ pp.\ \bibinfo {pages} {1--50}\BibitemShut {NoStop}%
\bibitem [{\citenamefont {Weinberg}(1995)}]{weinberg1995quantum}%
  \BibitemOpen
  \bibfield  {author} {\bibinfo {author} {\bibfnamefont {S.}~\bibnamefont
  {Weinberg}},\ }\href {https://books.google.com.br/books?id=doeDB3_WLvwC}
  {\emph {\bibinfo {title} {The Quantum Theory of Fields}}},\ \bibinfo {series}
  {Quantum Theory of Fields, Vol. 2: Modern Applications}\ No.\ \bibinfo
  {number} {v. 1}\ (\bibinfo  {publisher} {Cambridge University Press},\
  \bibinfo {year} {1995})\BibitemShut {NoStop}%
\bibitem [{\citenamefont {Polyakov}\ and\ \citenamefont
  {Wiegmann}(1983)}]{polyakov1983theory}%
  \BibitemOpen
  \bibfield  {author} {\bibinfo {author} {\bibfnamefont {A.}~\bibnamefont
  {Polyakov}}\ and\ \bibinfo {author} {\bibfnamefont {P.~B.}\ \bibnamefont
  {Wiegmann}},\ }\href@noop {} {\bibfield  {journal} {\bibinfo  {journal}
  {Physics Letters B}\ }\textbf {\bibinfo {volume} {131}},\ \bibinfo {pages}
  {121} (\bibinfo {year} {1983})}\BibitemShut {NoStop}%
\bibitem [{\citenamefont {Hooft}(1993)}]{hooft1993dimensional}%
  \BibitemOpen
  \bibfield  {author} {\bibinfo {author} {\bibfnamefont {G.}~\bibnamefont
  {Hooft}},\ }\href@noop {} {\bibfield  {journal} {\bibinfo  {journal} {arXiv
  preprint gr-qc/9310026}\ } (\bibinfo {year} {1993})}\BibitemShut {NoStop}%
\bibitem [{\citenamefont {Susskind}(1995)}]{susskind1995world}%
  \BibitemOpen
  \bibfield  {author} {\bibinfo {author} {\bibfnamefont {L.}~\bibnamefont
  {Susskind}},\ }\href@noop {} {\bibfield  {journal} {\bibinfo  {journal}
  {Journal of Mathematical Physics}\ }\textbf {\bibinfo {volume} {36}},\
  \bibinfo {pages} {6377} (\bibinfo {year} {1995})}\BibitemShut {NoStop}%
\bibitem [{\citenamefont {Donnay}\ \emph {et~al.}(2022)\citenamefont {Donnay},
  \citenamefont {Fiorucci}, \citenamefont {Herfray},\ and\ \citenamefont
  {Ruzziconi}}]{donnay2022carrollian}%
  \BibitemOpen
  \bibfield  {author} {\bibinfo {author} {\bibfnamefont {L.}~\bibnamefont
  {Donnay}}, \bibinfo {author} {\bibfnamefont {A.}~\bibnamefont {Fiorucci}},
  \bibinfo {author} {\bibfnamefont {Y.}~\bibnamefont {Herfray}},\ and\ \bibinfo
  {author} {\bibfnamefont {R.}~\bibnamefont {Ruzziconi}},\ }\href@noop {}
  {\bibfield  {journal} {\bibinfo  {journal} {Physical Review Letters}\
  }\textbf {\bibinfo {volume} {129}},\ \bibinfo {pages} {071602} (\bibinfo
  {year} {2022})}\BibitemShut {NoStop}%
\bibitem [{\citenamefont {Bagchi}, \citenamefont {Dhivakar},\ and\
  \citenamefont {Dutta}(2023)}]{bagchi2023ads}%
  \BibitemOpen
  \bibfield  {author} {\bibinfo {author} {\bibfnamefont {A.}~\bibnamefont
  {Bagchi}}, \bibinfo {author} {\bibfnamefont {P.}~\bibnamefont {Dhivakar}},\
  and\ \bibinfo {author} {\bibfnamefont {S.}~\bibnamefont {Dutta}},\
  }\href@noop {} {\bibfield  {journal} {\bibinfo  {journal} {Journal of High
  Energy Physics}\ }\textbf {\bibinfo {volume} {2023}},\ \bibinfo {pages} {1}
  (\bibinfo {year} {2023})}\BibitemShut {NoStop}%
\end{thebibliography}%

\end{document}